\documentclass[sigconf,nonacm,balance=false,review=false]{acmart}
\usepackage{amsmath,amssymb,amsfonts}
\usepackage{algorithm}
\usepackage{algpseudocode}
\usepackage{graphicx}
\usepackage{textcomp}
\usepackage{xcolor}
\usepackage{diagbox}
\usepackage{multirow}
\usepackage{url}
\usepackage[export]{adjustbox}
\usepackage[inline]{enumitem}
\usepackage{pifont}
\usepackage{lipsum}
\usepackage{wrapfig}
\usepackage{numprint}
\usepackage{calc}
\npthousandsep{\hspace{.875pt}}
\newcommand{\ybox}[3]{\centering\resizebox{#1\linewidth}{#2}{#3}}
\newcommand{\fybox}[3]{\ybox{#1}{#2}{\fbox{#3}}}

\algrenewcommand\textproc{}

\renewcommand{\checkmark}{\ding{52}}

\newcommand\tfun{\rightarrow}
\newcommand\fun{\mathrel{\ooalign{\hfil$\mapstochar\mkern18mu$\hfil\cr$\to$\cr}}}

\newcommand{\Chi}{{\mathcal{X}}}
\newcommand{\Tau}{\mathcal{T}}
\renewcommand{\L}{\mathcal{L}}
\renewcommand{\S}{\mathcal{S}}
\newcommand{\A}{\mathcal{A}}

\newcommand{\AP}{\mathcal{AP}}

\newcommand{\N}{\mathbb{N}}
\newcommand{\R}{\mathbb{R}}

\renewcommand{\U}{\mathcal{U}}\newcommand{\X}{{\bigcirc}}

\newcounter{savedenum}

\AtBeginDocument{\providecommand\longtitle{{Reinforcement Learning-Driven Test Generation for Android GUI Applications using Formal Specifications}}}

\makeatletter
\renewcommand{\ALG@beginalgorithmic}{\small}
\makeatother

\setcopyright{none}
\acmYear{2019}
\acmDOI{}

\acmConference[arXiv]{arXiv}{2019}{arXiv}
\acmBooktitle{arXiv}
\acmPrice{00.00}
\acmISBN{}

\begin{document}
\title[{\longtitle}]{\longtitle}

\author{Yavuz Koroglu}
\email{yavuz.koroglu@boun.edu.tr}
\orcid{0000-0001-9376-0698}
\affiliation{%
  \institution{Bogazici University}
  \city{Istanbul}
  \country{Turkey}
}

\author{Alper Sen}
\email{alper.sen@boun.edu.tr}
\orcid{0000-0002-5508-6484}
\affiliation{%
  \institution{Bogazici University}
  \city{Istanbul}
  \country{Turkey}
}

\renewcommand{\shortauthors}{Yavuz Koroglu and Alper Sen}

\begin{abstract}
There have been many studies on automated test generation for mobile Graphical User Interface (GUI) applications. These studies successfully demonstrate how to detect fatal exceptions and achieve high code and activity coverage with fully automated test generation engines. However, it is unclear how many GUI functions these engines manage to test. Furthermore, these engines implement only implicit test oracles. We propose Fully Automated Reinforcement LEArning-Driven Specification-Based Test Generator for Android (FARLEAD-Android). FARLEAD-Android accepts a GUI-level formal specification as a Linear-time Temporal Logic (LTL) formula. By dynamically executing the Application Under Test (AUT), it learns how to generate a test that satisfies the LTL formula using Reinforcement Learning (RL). The LTL formula does not just guide the test generation but also acts as a specified test oracle, enabling the developer to define automated test oracles for a wide variety of GUI functions by changing the formula. Our evaluation shows that FARLEAD-Android is more effective and achieves higher performance in generating tests for specified GUI functions than three known approaches, Random, Monkey, and QBEa. To the best of our knowledge, FARLEAD-Android is the first fully automated mobile GUI testing engine that uses formal specifications.
\end{abstract}




\maketitle

\section{Introduction}
\label{sec:intro}
Today, mobile GUI applications are ubiquitous, as there are more than 2.6 billion smartphone users worldwide \cite{Piejko:2016:DEVICEATLAS}. A recent survey shows that 78$\%$ of mobile GUI application users regularly encounter bugs that cause the GUI application to fail at performing some of its functions \cite{APPLAUSE}. Testing if the GUI application performs its intended functions correctly is essential for mitigating this problem.

More than 80$\%$ of the applications in the mobile market are Android GUI applications \cite{Gartner:2018} and there are 2 million Android GUI applications in Google Play \cite{GPLAYDATA}. Therefore, many studies propose fully automated test generation engines for Android GUI applications \cite{AzimNeamtiu:2013:OOPSLA,Anand+:2012:FSE,Moran+:2016:ICST,Cao+:2018:Internetware,Li+:2017:ICSE-C,Machiry+:2013:FSE,Mahmood+:2014:FSE,Yan+:2018:ISSTA,Amalfitano+:2015:IEEE,Google:MONKEY,Linares-Vasquez+:2015:MSR,Yang+:2013:FASE,Hao+:2014:MOBISYS,Koroglu+:2018:ICST,Mao+:2016:ISSTA,Su+:2017:FSE,Choi+:2013:OOPSLA,Choi:SWIFTHAND2,KorogluSen:2018:FASE,Mirzaei+:2016:ICSE,Cao+:2019:COMPSAC,Eler+:2018:ICST,Zaeem+:2014:ICST,Liu+:2014:IEEE}. These engines are adept at detecting fatal exceptions and achieving high code and activity coverage. However, it is unclear how many of the GUI functions they manage to test. In practice, an automated test generation engine may achieve high coverage but fail to test many essential GUI functions. Furthermore, engines in \cite{AzimNeamtiu:2013:OOPSLA,Moran+:2016:ICST,Cao+:2018:Internetware,Li+:2017:ICSE-C,Machiry+:2013:FSE,Yan+:2018:ISSTA,Amalfitano+:2015:IEEE,Koroglu+:2018:ICST,Mao+:2016:ISSTA,Su+:2017:FSE,KorogluSen:2018:FASE} only check fatal exceptions, \cite{Mirzaei+:2016:ICSE,Cao+:2019:COMPSAC,Anand+:2012:FSE,Mahmood+:2014:FSE,Google:MONKEY,Linares-Vasquez+:2015:MSR,Yang+:2013:FASE,Hao+:2014:MOBISYS,Choi+:2013:OOPSLA,Choi:SWIFTHAND2} only target coverage, and \cite{Eler+:2018:ICST,Zaeem+:2014:ICST,Liu+:2014:IEEE} only focus on accessibility issues, energy bugs, and app-agnostic problems, respectively. All these approaches ignore other possible functional bugs. 

To test if a GUI function works correctly, we must have an automated test oracle. All the test oracles mentioned until now (fatal exceptions, accessibility issues, app-agnostic problems, and energy bugs) are implicit \cite{Barr+:2014:TSE}, meaning that they are implemented using some implied assumptions and conclusions. Since there can be many different GUI functions, implementing implicit oracles for every GUI function is impractical. Therefore, we must use specified test oracles where each test oracle is associated with a formal specification. This way, the problem reduces to checking if the generated test satisfies the formal specification. However, checking if a test satisfies a formal specification is not enough to fully automate the test generation process. We must also develop an efficient method to generate the test. 

In this study, we use Reinforcement Learning (RL) to generate a test that satisfies a given GUI specification. RL is a semi-supervised machine learning methodology that has driven impressive advances in artificial intelligence in recent years, exceeding human performance in domains ranging from resource management \cite{Mao+:ACM:2016}, traffic light control \cite{Arel+:2010:IET}, playing sophisticated games such as chess \cite{Silver+:2017:arXiv} and atari \cite{Mnih+:2013:arXiv}, to chemistry \cite{Zhou+:ACS:2017}. In RL, an RL agent dynamically learns to perform its task by trial-and-error.  After every action, the RL agent receives an immediate reward from the environment. This reward can be positive, negative, or zero, meaning that the last decision of the RL agent was good, bad, or neutral, respectively. Decisions made according to the RL agent's experience is said to follow the agent's policy. After enough iterations, the RL agent becomes proficient in its task. At this point, the RL agent is said to have converged to its optimal policy. Upon convergence, the RL agent is said to have minimized its expected number of bad decisions in the future. The RL agent requires no prepared training data, which decreases the manual effort spent preparing it. Therefore, RL is attractive amongst many machine learning methods. 

Through dynamic execution, the RL agent learns from positive and negative rewards, on-the-fly for every action taken. Typically, an RL agent is trained to keep getting positive rewards and avoid the negative ones, indefinitely. Instead, our goal is to generate one satisfying test for a specified test oracle and terminate, which requires much less training than typical RL use cases. We exploit this fact to develop a test generator with low execution costs, which is crucial for dynamic execution tools.

\begin{figure}
\includegraphics[width=.68\linewidth]{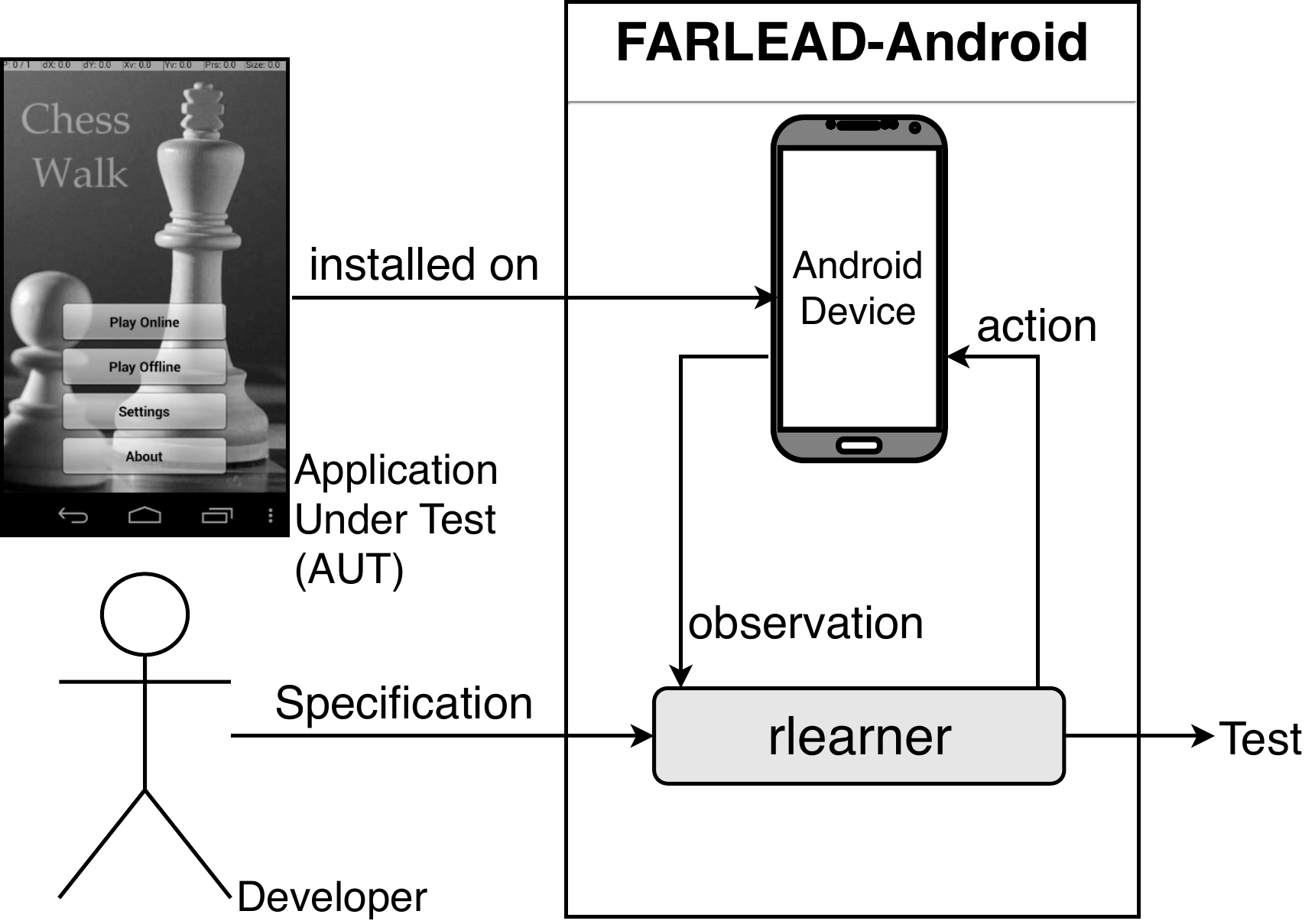}
\caption{FARLEAD-Android Overview}\label{fig:overview}
\Description{FARLEAD-Android Overview}
\end{figure}

We call our test generator Fully Automated Reinforcement LEAr\-n\-ing-Driven Specification-Based Test Generator for Android (FAR\-LEAD-Andro\-id) and show its overview in Figure~\ref{fig:overview}. First, FARLEAD-Android requires an Android Device where the Application Under Test (AUT) is installed. Second, the developer provides a Linear-time Temporal Logic (LTL) formula as a formal specification. \emph{rlearner} takes this LTL formula as input and performs RL by sending replayable actions to the Android Device and then receiving observations from the Android Device. The replayable action sequence generated during this process is called a test. \emph{rlearner} checks if the generated test satisfies the LTL formula and accordingly calculates a reward after every action. Finally, as soon as \emph{rlearner} finds a satisfying test, it outputs the test and terminates.

FARLEAD-Android generates tests through dynamic execution. Hence, there is an execution cost for every action. To reduce the execution cost as much as possible, we implement several improvements which we describe in the following paragraphs.

Traditional RL approaches get positive rewards only when the objective is satisfied. However, since we are interested in test generation we terminate as soon as we find a satisfying test, which restricts a typical RL agent to learning only from negative rewards. We propose LTL-based Reward Shaping, which generates positive intermediate rewards. To the best of our knowledge, ours is the first work that combines LTL and Reward Shaping.

Actions enabled in a GUI state directly affects the RL performance. If there are too many actions enabled, it takes a long time for FARLEAD-Android to generate a satisfying test. We demonstrate that if the LTL formula specifies properties about not just the observations but also actions, we can predict if the LTL formula is going to be satisfied before action execution, saving time. We call this improvement Reward Prediction.

We argue that the satisfaction of an LTL formula may not just depend on the current GUI state, but also on previous GUI states. In that case, the same state-action pair may get conflicting rewards and confuse a typical RL agent. To address this issue, we propose to encode GUI states as a sequence of action-state pairs which we call tails. Then, we encode actions as tail-action pairs which we call decisions. We put a maximum tail length  to avoid state explosion. 

It is common to start RL with indifference to all available decisions. This indifference leads to the random exploration of all new GUI states. We propose to use the knowledge about previously encountered actions in new GUI states by learning what we call stateless values. We call this novel stateless learning improvement as Action Label Learning.

In short, our main contributions in this paper are as follows.

\begin{enumerate}

\item To the best of our knowledge, FARLEAD-Android is the first fully automated mobile GUI test generation engine that uses formal specifications.

\item We implement four novel improvements in RL; LTL-based Reward Shaping, Reward Prediction, Tails/Decisions, and Action Label Learning.

\item We evaluate FARLEAD-Android via experiments on two applications from F-Droid. We show that our approach is more effective and achieves higher performance in generating satisfying tests than three known test generation approaches, namely Monkey, Random, and QBEa.

\end{enumerate}

We give background information, describe our approach, provide an example, make evaluations, explain related work, discuss issues, and conclude in Sections~\ref{sec:background}-\ref{sec:conclusion}, respectively.

\section{Background}
\label{sec:background}
We now provide information on Android Devices and Linear-time Temporal Logic (LTL) in Sections~\ref{sec:background:subsec:android} and \ref{sec:background:subsec:ltl}, respectively.

\subsection{Android Devices}
\label{sec:background:subsec:android}
We analyze an Android Device as a system which takes GUI actions as input and produces GUI states as output where we denote the set of all GUI states and GUI actions as $S$ and $A$, respectively. A GUI state or state $s \in S$ is the set of observed attributes about 
\begin{enumerate*}
\item the active package, \item the focused activity, \item widgets on the screen, and \item contextual properties such as the screen being on or off.
\end{enumerate*}
 In any GUI state, the user interacts with the Android Device using click and other gestures, which we collectively call GUI actions or actions. A GUI action $a \in A$ is either
\begin{enumerate*} 
\item widget-dependent such as \emph{click} and \emph{text} or \item widget-independent such as \emph{back} which presses the hardware back button. 
\end{enumerate*}
Due to widget-dependency, every GUI state has a different set of enabled actions $A_s \subseteq A$. Note that a GUI action may have parameters. For example, a \emph{click} requires two coordinates.

Initially, we assume that the Android Device is in a don't care state and we denote this GUI state as $\varnothing \in S$. In a don't care state, we allow only \emph{reinitialize} actions. A \emph{reinitialize} action reinstalls the AUT and starts one of its launchable activities given as a parameter. We do not allow any \emph{reinitialize} action in any other GUI state. We describe how we decide to restart the AUT at a GUI state other than $\varnothing \in S$ in Section 3.

A dynamic execution trace, or just trace, is a finite sequence of action-state pairs where the first action is a \emph{reinitialize} action. We denote it as $t = a_0 s_0 a_1 s_1 \ldots a_{|t|-1} s_{|t|-1}$ where $|t| \in \N$ is the trace length. Note that in this definition, we omit $\varnothing \in S$ which is always the initial state. A test $ts = a_0 a_1 \ldots a_{|t|-1}$ is the finite action sequence that generates the GUI states in the trace. 

In Figure 1, we make three kinds of observations on the Android Device. We observe 
\begin{enumerate*}
\item the set of currently enabled actions $A_{s_k} \subseteq A$, 
\item the resulting state $s_{k+1} \in S$ after executing the selected action $a_{k+1} \in A_{s_k}$, and 
\item a set of atomic propositions that currently evaluate to true ($\top$).
\end{enumerate*}
An atomic proposition that evaluates to true is called a label and denoted as $l \in \AP$ where $\AP$ is the set of all possible atomic propositions. A label $l \in \AP$ is an action label if it is an observation on a GUI action $a \in A$. Otherwise, it is an observation on a state $s \in S$, so it is a state label. 
The set of all observed state and action labels are called state and action labelings, $\L(s) \in 2^\AP$ and $\L(a) \in 2^\AP$, respectively.

Formally, we represent the observations as an Observation Triplet $\Omega(\Tau,\Chi,\L)$ where 
\begin{enumerate*}
\item $\Tau : S \tfun 2^A$ is a function which returns the set of enabled actions $A_s \subseteq A$ in the current state $s \in S$, \item $\Chi : (S \times A) \tfun S$ is a function that returns the next state after executing the action, and \item $\L : (S \cup A) \tfun 2^\AP$ is a labeling function for GUI states and actions.
\end{enumerate*}
The Observation Triplet $\Omega$ defines the necessary conditions to perform RL and the interface between the Android Device and \emph{rlearner}.

\subsection{Linear-time Temporal Logic (LTL)}
\label{sec:background:subsec:ltl}
\begin{table}
\caption{Pointwise LTL Semantics}\label{tbl:pointwise}
\fybox{.75}{!}{\begin{tabular}{rlll}
$(t, k)$ & \hspace*{-6pt}$\models \top$ \\
$(t, k)$ & \hspace*{-6pt}$\models p$ & iff & $p \in \L(a_k) \cup \L(s_k)$ \\
$(t, k)$ & \hspace*{-6pt}$\models \neg \phi$ & iff & $(t, k) \not\models \phi$ \\
$(t, k)$ & \hspace*{-6pt}$\models \phi \wedge \phi'$ & iff & $(t, k) \models \phi$ and $(t, k) \models \phi'$ \\
$(t, k)$ & \hspace*{-6pt}$\models \X \phi$ & iff & $(t, k+1) \models \phi$ \\
$(t, k)$ & \hspace*{-6pt}$\models \phi \U \phi'$ & iff & $\exists j \in \N, k \leq j < |t|, (t, j) \models \phi',$ \\
			&	& and & $\forall i \in \N, [k \leq i < j \rightarrow (t, i) \models \phi]$  \\  
$t$ & \hspace*{-6pt}$\models \phi$ & iff & $t, 0 \models \phi$ \\
\end{tabular}}
\end{table}

Equation~\eqref{eqn:ltlsyntax} defines the syntax of an LTL formula $\phi$ where $p \in \AP$ is an atomic proposition.
\begin{equation}
\phi := \top | p | \neg \phi | \phi \wedge \phi | \X \phi | \phi \U \phi \label{eqn:ltlsyntax}
\end{equation}

We interpret $\phi$ over a (finite) trace $t = a_0 s_0 a_1 s_1 \ldots$ $a_{|t|-1} s_{|t|-1}$ using the pointwise semantics in Table~\ref{tbl:pointwise}. A test $ts \in A^*$ is satisfying if and only if its execution trace $t \in (A \times S)^*$ satisfies the formula $\phi$.

\section{Methodology}
\label{sec:method}
\begin{algorithm}[!t]
\caption{A General RL Framework for \emph{rlearner}}\label{alg:rlearner}
\begin{algorithmic}[1]
\Require 
\Statex $\Omega = (\Tau, \Chi, \L)$ is the observation triplet
\Statex $E, K \in \N$ are the maximum number of episodes and steps, respectively
\Ensure
\Statex $ts \in \bigcup_{k=0}^K A^k$ is a satisfying test or $i > E$
\Statex \vspace*{-.5\baselineskip}
\State initializeVariables()\label{line:initVars:alg:rlearner}
\State $i \gets 1$ \Comment{Episode index}\label{line:epindex:alg:rlearner}
\Repeat
\State $k \gets 0$ \Comment{Position index}\label{line:posindex:alg:rlearner}
\State $s_{-1} \gets \varnothing$ \Comment{Start from a \emph{don't care} state}\label{line:startstate:alg:rlearner}
\State initializeEpisode()\label{line:initEpisode:alg:rlearner}
\Repeat
\State $a_k \gets \text{decideNextAction}(s_{k-1}, \Tau(s_{k-1}))$\label{line:decide:alg:rlearner}
\State $s_k \gets \Chi(s_{k-1}, a_k)$\label{line:execute:alg:rlearner}
\State $r \gets \text{calculateImmediateReward}(k, \L(a_k) \cup \L(s_{k}))$\label{line:reward:alg:rlearner}
\State learn$(s_{k-1}, a_k, r, s_k)$\label{line:learn:alg:rlearner}
\State $k \gets k + 1$\label{line:incrementposindex:alg:rlearner}
\Until{$r = 1$ \textbf{or} $r = -1$ \textbf{or} $k \geq K$}\label{line:inneruntil:alg:rlearner}
\State makePostEpisodeUpdates()\label{line:postep:alg:rlearner}
\State $ts \gets a_0 \ldots a_{k-1}$\label{line:test:alg:rlearner}
\State $i \gets i + 1$\label{line:incrementepindex:alg:rlearner}
\Until{$r = 1$ or $i > E$}\label{line:outeruntil:alg:rlearner}
\end{algorithmic}
\end{algorithm}

Reinforcement Learning (RL) \cite{SuttonBarto:1998:RL} is a semi-supervised machine learning methodology. The main idea of RL is to calculate a reward at every state, so previous rewards drive future decisions.

In Figure~\ref{fig:overview}, \emph{rlearner} is responsible for calculating rewards from observations and generate a satisfying test. We describe a general RL framework for \emph{rlearner} in Algorithm~\ref{alg:rlearner}, with six procedures explained in later sections. Algorithm~\ref{alg:rlearner} has three requirements. First, it requires the Observation Triplet $\Omega$, which is the interface between the Android Device and \emph{rlearner}. Second, it requires a maximum number of episodes $E$, where $E$ is equal to the maximum number of tests generated before termination. If $E$ is too small, \emph{rlearner} may often terminate without finding a satisfying test. If $E$ is too large, \emph{rlearner} may spend too much time, especially over unsatisfiable specifications. The third requirement is the maximum number of steps $K$. At every step, \emph{rlearner} executes exactly one GUI action, so $K$ is equal to the maximum trace length. If $K$ is too small, \emph{rlearner} may not find any satisfying test. If $K$ is too large, \emph{rlearner} may generate unnecessarily long tests.

We initialize the variables required for RL in Line~\ref{line:initVars:alg:rlearner} of Algorithm~\ref{alg:rlearner}. We also initialize an episode and a position index in Lines~\ref{line:epindex:alg:rlearner} and \ref{line:posindex:alg:rlearner}, respectively.  Every episode starts from a don't care state $s_{-1} = \varnothing$, as shown in Line~\ref{line:startstate:alg:rlearner}. Before every episode, we update RL-related variables in Line \ref{line:initEpisode:alg:rlearner}. We divide an episode into at most $K$ number of steps. At every step, 
\begin{enumerate*}
\item we decide an action $a_k \in A_{s_{k-1}}$ in Line \ref{line:decide:alg:rlearner}, 
\item execute the action $a_k$ and obtain the resulting state $s_k \in S$ in Line \ref{line:execute:alg:rlearner}, 
\item calculate a reward $r \in [-1,1]$ from labelings of the current position $\L(a_k)$ and $\L(s_k)$ in Line~\ref{line:reward:alg:rlearner}, 
\item learn from the immediate reward $r$ in Line~\ref{line:learn:alg:rlearner}, and finally 
\item increment the position index $k$ in Line~\ref{line:incrementposindex:alg:rlearner}.
\end{enumerate*}
If the immediate reward $r$ is equal to $1$, we understand that the test satisfies the given specification. If the immediate reward $r$ is equal to $-1$, we understand the latest action $a_k$ made it impossible to satisfy the given specification, no matter what actions we choose from now on. We say that a test $ts$ is a dead-end if one of its actions gets a reward of $-1$. We end the episode prematurely if we get either $1$ or $-1$ as a reward in Line~\ref{line:inneruntil:alg:rlearner}. Otherwise, we continue until we generate $K$ steps. After every episode, we update RL-related variables in Line~\ref{line:postep:alg:rlearner}. The test $ts$ is the action sequence generated during the last episode, as shown in Line~\ref{line:test:alg:rlearner}. We increment the episode index $i$ in Line~\ref{line:incrementepindex:alg:rlearner}. We continue to a new episode until either the immediate reward $r$ is equal to $1$, which means that the test $ts$ is satisfying, or we reach the maximum number of episodes $E$.

We now explain FARLEAD-Android by implementing the procedures in Algorithm~\ref{alg:rlearner}, through Sections~\ref{sec:method:subsec:policy}-\ref{sec:method:subsec:learning}.

\subsection{Policy}
\label{sec:method:subsec:policy}
\begin{algorithm}[!t]
\caption{$\epsilon$-greedy Softmax Policy with Temperature}\label{alg:policy}
\begin{algorithmic}[1]
\Procedure{decideNextAction}{$s \in S, A_s \subseteq A$}\label{line:decide:alg:policy}
\ForAll{$a \in A_s$}
\State $P(a|s) \gets (1-\epsilon) \frac{\exp\left((Q_1(s, a) + Q_2(s, a))/2T\right)}{\sum_{a' \in A_s} \exp\left((Q_1(s, a') + Q_2(s, a'))/2T\right)} + \epsilon\frac{1}{|A_s|}$
\EndFor
\State \textbf{return} \textbf{random} $a \in A_s$ \textbf{with probability} $P(a|s)$
\EndProcedure
\end{algorithmic}
\end{algorithm}

A policy is a procedure which decides on a next action $a$ from the set of currently enabled actions $A_s$ in the current state $s \in S$. Traditionally, the $\epsilon$-greedy and the softmax policies are the most common policies in RL \cite[p.525]{Alpaydin:2014:ML}. We unify the common softmax and $\epsilon$-greedy policies in Algorithm~\ref{alg:policy}. This policy implements the procedure given in Line~\ref{line:decide:alg:rlearner} of Algorithm~\ref{alg:rlearner}. In the policy, $\epsilon \in [0,1]$ is the probability of deciding the action, completely randomly, and $T \in \R$ denotes the temperature variable. High temperatures indicate low trust and vice versa. Intuitively, it makes sense to start with a high temperature and gradually decrease it (increase trust) as the learning continues. This approach is called annealing.  We describe quality functions, Q-functions in short, to explain our policy. At every step, RL learns from the immediate reward $r \in [-1,1]$ by updating a quality function $Q: (S \times A) \fun \R$. Using two Q-functions ($Q_1$ and $Q_2$) instead of one is an improvement known as Double Q-Learning. Double Q-Learning improves learning performance by reducing maximization bias (overestimation). Q-values of these Q-functions imply a probability distribution $P$, as shown in Line 3. We sample a GUI action from this probability distribution in Line 5.

\paragraph{Reward Prediction.} It is trivial to see that the number of enabled actions $|A_s|$ directly affects learning performance. So, it is essential to keep $|A_s|$ as small as possible. Therefore, we modify Algorithm~\ref{alg:rlearner} by adding three operations immediately before Line~\ref{line:decide:alg:rlearner}. 
First, thanks to the observation triplet $\Omega = (\Tau, \Chi, \L)$, we know the action labeling $\L(a)$ before executing the action $a$. We use this fact to eliminate all GUI actions from $\Tau(s_{k-1})$ that make the specification false ($\neg \top$) from the set of enabled actions. 
Second, if $|A_s|$ becomes zero, we understand that all actions in the current state lead to dead-ends, so we conclude that the previous state-action pair leads to a dead-end. Therefore, we terminate the episode with a reward of $-1$ to the previous state-action pair $(s_{k-2}, a_{k-1})$. 
Finally, if $|A_s|$ was not zero at the second step and we find one action that makes the specification true ($\top$), we immediately take that action and terminate FARLEAD-Android, since we have found a satisfying test. If we cannot find such an action, we continue from Line~\ref{line:decide:alg:rlearner}. We call this new improvement Reward Prediction. 

\subsection{Rewards}
\label{sec:method:subsec:reward}
\begin{algorithm}[!t]
\caption{LTL-based Reward Calculation with Reward Shaping}\label{alg:reward}
\begin{algorithmic}[1]
\Procedure{calculateImmediateReward}{$k \in \N, L \in 2^\AP$}
\State $\phi_{k+1} \gets \text{projection}(\phi_{k}, L)$\label{line:projection:alg:reward}
\State \textbf{return} $\left\lbrace \begin{matrix} 1 & \phi_{k+1} = \top \\ -1 & \phi_{k+1} = \neg \top \\ \frac{|N(\phi_{k+1}) - N(\phi_{k})|}{N(\phi_{k+1}) + N(\phi_{k})} RS & \text{otherwise} \end{matrix} \right.$\label{line:shaping:alg:reward}
\EndProcedure
\Statex \vspace*{-.5\baselineskip}
\Procedure{projection}{$\phi\text{ in LTL}, L \in 2^\AP$}
\State \textbf{return} advance(restrict(expand($\phi$), $L$))
\EndProcedure
\Statex \vspace*{-.5\baselineskip}
\Procedure{expand}{$\phi$ in LTL}
\State \textbf{return} $\left\lbrace \begin{matrix} \neg\text{expand}(\phi') & \phi = \neg \phi' \\ \text{expand}(\phi') \wedge \text{expand}(\phi'') & \phi = \phi' \wedge \phi'' \\ \neg (\neg \text{expand}(\phi'') \wedge \neg (\text{expand}(\phi') \wedge \X \phi) & \phi = \phi' \U \phi'' \\ \phi & \text{otherwise} \end{matrix} \right.$
\EndProcedure
\Statex \vspace*{-.5\baselineskip}
\Procedure{restrict}{$\phi\text{ in LTL}, L \in 2^\AP$}
\State \textbf{return} $\left\lbrace \begin{matrix} \top & \phi = \top\textbf{ or }\phi \in L \\ \neg \top & \phi \in \AP\textbf{ but }\phi \not\in L \\ \neg\text{restrict}(\phi') & \phi = \neg \phi' \\ \text{restrict}(\phi') \wedge \text{restrict}(\phi'') & \phi = \phi' \wedge \phi'' \\ \phi & \text{otherwise} \end{matrix} \right.$
\EndProcedure
\Statex \vspace*{-.5\baselineskip}
\Procedure{advance}{$\phi$ in LTL}
\State \textbf{return} $\left\lbrace \begin{matrix} \text{advance}(\phi') & \phi = \neg \phi' \\ \text{advance}(\phi') \wedge \text{advance}(\phi'') & \phi = \phi' \wedge \phi'' \\ \phi' & \phi = \X \phi' \\ \phi & \text{otherwise} \end{matrix} \right.$
\EndProcedure
\end{algorithmic}
\end{algorithm}

The immediate reward is typically either one, zero, or minus one, depending on whether the RL agent achieved, not yet achieved, or cannot achieve its objective anymore in the current episode, respectively. The fact that FARLEAD-Android terminates after receiving the positive reward restricts the typical agent to learn only from negative rewards.

\paragraph{Reward Shaping.}
We develop an LTL-based {\em Reward Shaping} approach that enables us to produce intermediate rewards between zero and one in Algorithm~\ref{alg:reward}, so FARLEAD-Android can recieve positive rewards before termination. This algorithm implements the procedure in Line~\ref{line:reward:alg:rlearner} of Algorithm~\ref{alg:rlearner}.

First, in Line~\ref{line:projection:alg:reward}, we obtain the LTL formula $\phi_{k+1}$ from the LTL formula $\phi_k$ by using the labeling $L \in 2^\AP$. We describe how we get this new formula with an example in Section~\ref{sec:example}. We return one or minus one if the LTL formula $\phi_{k+1}$ is true or false, respectively. Otherwise, we return a reward between zero and one, depending on a distance metric between $\phi_k$ and $\phi_{k+1}$. We measure the distance using a function $N$ that returns the number of atomic propositions in the formula. The distance reflects the amount of change between two formulae and is zero if there is no change. As a final note, $RS$ is an enabling parameter that takes either one or zero to enable or disable the Reward Shaping, respectively.

We now explain the projection procedure. We divide this procedure into three sub-procedures, 
\begin{enumerate*}
\item \emph{expand},
\item \emph{restrict}, and
\item \emph{advance}.
\end{enumerate*}
In \emph{expand}, we apply the expansion law to all $\U$ operators in the formula. We present the expansion law in Equation~\eqref{eqn:expansionlaw}. In \emph{restrict}, we substitute the atomic propositions with trues and falses according to the labeling $L$. Finally, in \emph{advance}, we eliminate one round of $\X$ operators.
\begin{equation}
\phi' \U \phi'' = \neg (\neg \phi'' \wedge \neg (\phi' \wedge \X (\phi' \U \phi''))\label{eqn:expansionlaw}
\end{equation}

\subsection{Learning}
\label{sec:method:subsec:learning}
\begin{algorithm}[!t]
\caption{Learning Procedures}\label{alg:myopiclearning}
\begin{algorithmic}[1]
\Procedure{initializeVariables}{}
\State $(Q_1, Q_2, T, \epsilon, \eta) \gets (Q_0, Q_0, T_0, \epsilon_0, \eta_0)$\label{line:allinit:alg:myope}
\EndProcedure
\Statex \vspace*{-.5\baselineskip}
\Procedure{initializeEpisode}{}
\State $e \gets Q_0$\label{line:epinit:alg:myope}
\EndProcedure
\Statex \vspace*{-.5\baselineskip}
\Procedure{learn}{$s \in S, a \in A_s, r \in [-1,1], s' \in S$}\label{line:asiloglan:alg:myope}
\State $\delta \gets r - Q_1(s, a)$\Comment{Myopic Update Equation}\label{line:delta:alg:myope}
\State $e(s, a) \gets e(s, a) + 1$\label{line:estart:alg:myope}
\ForAll{$(s, a) \in S \times A$ \textbf{s.t.} $e(s, a) \geq e_{\min}$}
\State $Q_1(s, a) \gets \min( \max( Q_1(s, a) + \eta \delta e(s,a), -\rho ), \rho )$ \label{line:backpropagation:alg:myope}
\State $Q_2(s, a) \gets (1 - \alpha) Q_1(s, a) + \alpha Q_2(s, a)$ \label{line:double:alg:myope}
\State $e(s, a) \gets \lambda e(s, a)$ \label{line:ediscount:alg:myope}
\EndFor\label{line:eend:alg:myope}
\State $(Q_1, Q_2) \gets (Q_2, Q_1)$ \textbf{with probability} $.5$\label{line:swap:alg:myope}
\EndProcedure
\Statex \vspace*{-.5\baselineskip}
\Procedure{makePostEpisodeUpdates}{}
\State $(T,\hspace*{-1pt} \epsilon,\hspace*{-1pt} \eta)\hspace*{-2pt}\gets\hspace*{-2pt}(\max(T - \Delta T,\hspace*{-1pt} T_{\min}), \max(\epsilon_u \epsilon,\hspace*{-1pt} \epsilon_{\min}), \max(\eta_u \eta,\hspace*{-1pt} \eta_{\min}))$\label{line:final:alg:myope}
\EndProcedure
\end{algorithmic}
\end{algorithm}

Traditionally, RL stores the quality function $Q: (S \times A) \fun \R$ as a look-up table. Whenever it executes an action $a \in A$ at a state $s \in S$, it adds an update value $\delta \in \R$ multiplied by a learning rate $\eta \in \R$ to the Q-value $Q(s, a)$. The update value $\delta$ commonly depends on three terms, 
\begin{enumerate*}
\item the immediate reward, 
\item the previous Q-value of the current state-action pair, and 
\item a Q-value for a next state-action pair.
\end{enumerate*}
The third term allows future Q-values to backpropagate one step at a time. Instead of the third term, an {\em eligibility trace} $e: (S \times A) \fun \R$, which is an improvement used in RL,  allows for faster backpropagation. The main idea of using the eligibility trace is to remember previous state-action pairs and update all of them at once, directly backpropagating multiple steps at a time.

We describe all the remaining procedures of Algorithm~\ref{alg:rlearner} in Algorithm~\ref{alg:myopiclearning}. We initialize five variables in Line~\ref{line:allinit:alg:myope}, 
\begin{enumerate*}
\item the first Q-function $Q_1$, 
\item the second Q-function $Q_2$, 
\item the temperature $T$, 
\item the random decision probability $\epsilon$, and 
\item the learning rate $\eta$.
\end{enumerate*}
Note that $Q_0: (S \times A) \fun \lbrace 0 \rbrace$ is the initial Q-function where every state-action pair maps to zero. Before every episode, we initialize an eligibility trace $e$ to all zeros in Line~\ref{line:epinit:alg:myope}. 

The procedure in Line~\ref{line:asiloglan:alg:myope} requires four parameters, 
\begin{enumerate*}
\item the GUI state $s \in S$, 
\item the executed action $a \in A_s$, 
\item the immediate reward $r \in [-1,1]$, and 
\item the resulting state $s' \in S$. 
\end{enumerate*}
We give our update equation in Line~\ref{line:delta:alg:myope}. Our update equation does not take any future Q-values into account, so it does not use the resulting state $s'$. This is why our approach is also called {\em myopic learning}.

We ensure that the current state-action pair has non-zero eligibility in Line~\ref{line:estart:alg:myope}. We perform the update on all eligible state-action pairs by adding $\delta$ multiplied by $\eta$ and $e(s,a)$ to the Q-value $Q(s,a)$ in Line~\ref{line:backpropagation:alg:myope}. We force all Q-values to be between $[-\rho,\rho]$. This idea is known as {\em vigilance} in adaptive resonance theory, and it prevents the accumulation of too large Q-values. Our experience shows that too large Q-values may prevent further learning. $\alpha$ is the doubleness ratio. If it is zero, we ensure that $Q_2 = Q_1$ in Line~\ref{line:double:alg:myope}, and therefore Double-Q is disabled. If it is one, $Q_2$ remains completely disjoint from $Q_1$, and therefore Double-Q is enabled. We can also take $\alpha = .5$ to perform mixed Double-Q learning. We ensure that the eligibility trace updates closer state-action pairs more by multiplying it with an eligibility discount rate $\lambda$ in Line~\ref{line:ediscount:alg:myope}. We swap the two Q-functions half of the time to perform Double-Q learning. Note that if $\alpha = 0$, $Q_2 = Q_1$ and therefore this swap becomes redundant.

Finally, after every episode, we decrease 
\begin{enumerate*}
\item the temperature $T$ to perform annealing by a stepwise temperature difference $\Delta T$, 
\item the random decision probability $\epsilon$ by an update factor $\epsilon_u$ to follow the traditional $\epsilon$-greedy strategy, and 
\item the learning rate $\eta$ by another update factor $\eta_u$ to prevent forgetting previous Q-values
\end{enumerate*}
in Line~\ref{line:final:alg:myope}.

\paragraph{Action Label Learning.} A Q-value $Q(s, a)$ encodes information about only a single state-action pair, and therefore it is useful only if that state is the current state. Action Label Learning allows us to use  Q-values in new GUI states. The main idea is to learn stateless Q-values $Q_1^\AP: 2^\AP \rightarrow \R$ and $Q_2^\AP: 2^\AP \rightarrow \R$ for action labelings immediately before Line~\ref{line:backpropagation:alg:myope} of Algorithm~\ref{alg:myopiclearning}. Whenever we are going to observe a new GUI state $s \in S$ in Line~\ref{line:delta:alg:myope}, we update the Q-value as $Q_1(s, a) \leftarrow Q_1^\AP(\L(a))$ immediately before that line. Immediately before Line~\ref{line:double:alg:myope}, we execute $Q_2^\AP(\L(a)) \leftarrow (1 - \alpha) Q_1^\AP(\L(a)) + \alpha Q_2^\AP(\L(a))$. In Line~\ref{line:swap:alg:myope}, we also swap $Q_1^\AP$ and $Q_2^\AP$ with $.5$ probability.

\paragraph{Tails/Decisions.} So far, we assumed that we always produce the same immediate reward at a state regardless of how we reached that particular state, which is not always true. We propose the Tails/Decisions approach to mitigate this issue. In this approach, we replace the set of states $S$ and the set of actions $A$ with a finite set of tails $\S = \bigcup^h_{i = 0} (A \times S)^i$ where $h \in \N$ is the maximum tail length and a set of decisions $\A = \S \times A$, respectively. This way, we can learn different Q-values for the same state-action pair with different paths. Note that $h$ must be small to avoid state explosion.

\section{Example}
\label{sec:example}
We now describe how FARLEAD-Android works with a small example. Our Application Under Test (AUT) for this example is ChessWalk, which is a chess game. The GUI function that we are testing for is the ability to go from MainActivity to AboutActivity and then return to MainActivity. We specify this GUI function as $\phi_0 = \X (p \U (q \wedge \X (q \U p)))$, where $p$ and $q$ are true if and only if the current activity matches with the word Main and the word About, respectively. According to $\phi_0$, the activity of the second GUI state must be MainActivity until it is AboutActivity and then it must be AboutActivity until it is MainActivity. Note that we impose this condition on the second GUI state and not the first because the first GUI state is a don't care state.

\begin{figure}
\begin{scriptsize}
\raisebox{-.5\height}{\includegraphics[frame,width=.25\linewidth]{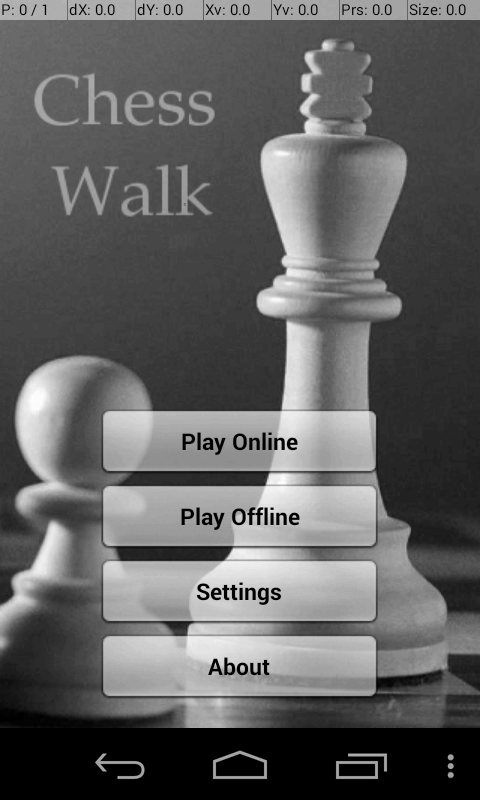}}$\;\xrightarrow{\begin{matrix}\text{pause}\\\text{resume}\end{matrix}}$
\raisebox{-.5\height}{\includegraphics[frame,width=.25\linewidth]{figures/MainGray.png}}$\;\xrightarrow{\;\;\text{back}\;\;}$
\raisebox{-.5\height}{\includegraphics[frame,width=.25\linewidth]{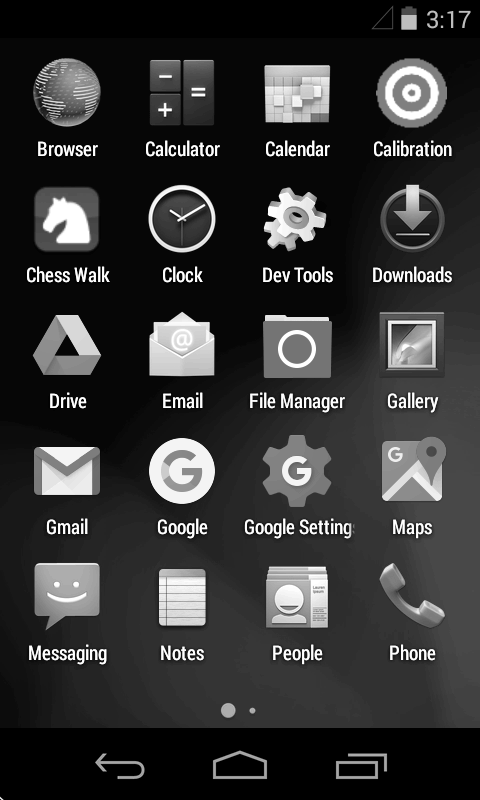}}
\caption{Episode i = 1}\label{fig:episode1}
\Description{Episode i = 1}
\end{scriptsize}
\end{figure}

\begin{figure}
\begin{scriptsize}
\raisebox{-.5\height}{\includegraphics[frame,width=.25\linewidth]{figures/MainGray.png}}$\;\xrightarrow{\begin{matrix}\text{click}\\\text{239 669}\end{matrix}}$
\raisebox{-.5\height}{\includegraphics[frame,width=.25\linewidth]{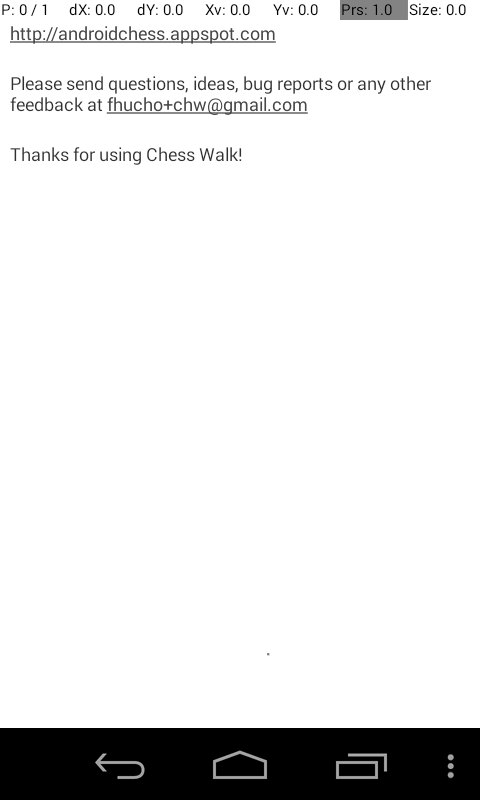}}$\;\xrightarrow{\begin{matrix}\text{click}\\\text{143 32}\end{matrix}}$
\raisebox{-.5\height}{\includegraphics[frame,width=.25\linewidth]{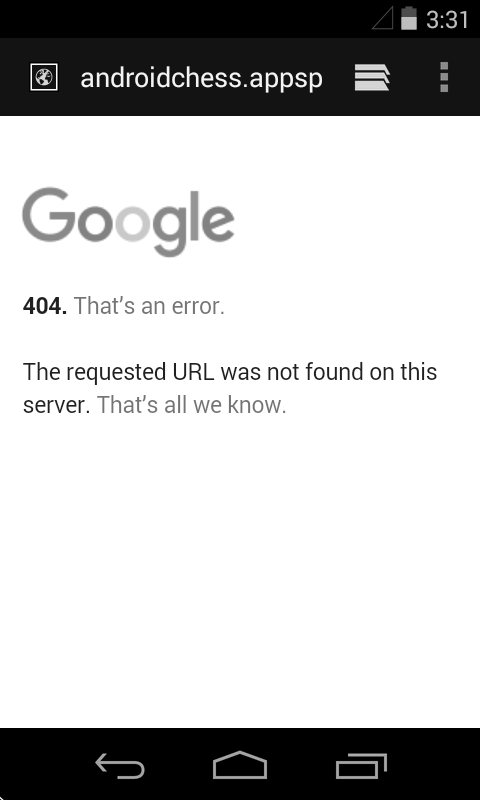}}
\caption{Episode i = 2}\label{fig:episode2}
\Description{Episode i = 2}
\end{scriptsize}
\end{figure}

\begin{figure}
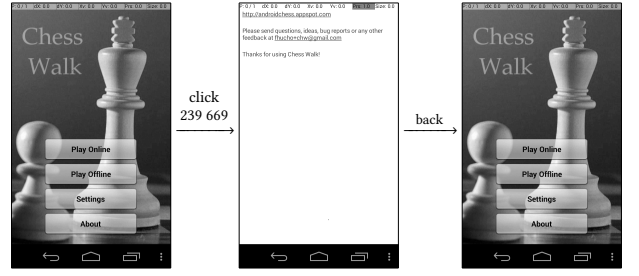

\begin{scriptsize}
\raisebox{-.5\height}{\includegraphics[frame,width=.25\linewidth]{figures/MainGray.png}}$\;\xrightarrow{\begin{matrix}\text{click}\\\text{239 669}\end{matrix}}$
\raisebox{-.5\height}{\includegraphics[frame,width=.25\linewidth]{figures/AboutGray.png}}$\;\xrightarrow{\;\;\text{back}\;\;}$
\raisebox{-.5\height}{\includegraphics[frame,width=.25\linewidth]{figures/MainGray.png}}
\caption{Episode i = 3}\label{fig:episode3}
\Description{Episode i = 3}
\end{scriptsize}
\end{figure}

In our example, FARLEAD-Android finds a satisfying test in three episodes. We provide the details of these episodes in Table~\ref{tbl:example} and give the screenshots for these episodes in Figures~\ref{fig:episode1}-\ref{fig:episode3}. For every episode, Table~\ref{tbl:example} shows the episode index $i \in \N$, the initial LTL formula $\phi_0$, and the position index $k \in \N$, the selected action $a_k \in A$, the labeling $L \in 2^\AP$ after executing $a_k \in A$, $\phi_k$ projected from the previous formula $\phi_{k-1}$, and the immediate reward $r \in [-1,1]$, for every step. The reinitialize action is the only choice at a don't care state, so every episode begins with $a_0 =\text{ reinitialize chesswalk MainActivity}$. The labeling after $a_0$ is $L = \lbrace p \rbrace$ because the resulting activity matches the word Main. We calculate the projection $\phi_1$ from $\phi_0$ as in Algorithm 3. Note that $\text{restrict}(\text{expand}(\phi_0), L)) = \phi_0$ because the first operator is $\X$, meaning that the formula states nothing about the current state and starts from the next state. We only advance $\phi_0$ one step by removing a single level of $\X$ operators and get $\phi_1 = p \U (q \wedge \X (q \U p))$. The number of atomic propositions in $\phi_0$ and $\phi_1$ are $N(\phi_0) = 4$ and $N(\phi_1) = 4$, respectively. Therefore, the immediate reward is $r = |N(\phi_1)-N(\phi_0)|/(N(\phi_1)+N(\phi_0)) = 0/8 = 0$. The zero reward means that $a_0$ is neither good nor bad for satisfying the formula.

In the first episode, FARLEAD-Android has no idea which action is going to satisfy the specification, so it chooses a random action $a_1 =\text{pauseresume}$. This action only pauses and then resumes the AUT, so we reach the same state, as shown in Figure~\ref{fig:episode1}. The labeling is again $L = \lbrace p \rbrace$. This time, we expand $\phi_1$ as in Equation~\eqref{eqn:expand1}.
\begin{equation}
\neg ( \neg (q \wedge \X (q \U p)) \wedge \neg (p \wedge \X (p \U (q \wedge \X (q \U p))))) \label{eqn:expand1}
\end{equation}

$L = \lbrace p \rbrace$ means $p = \top$ and $q = \neg \top$ in the current state. Hence, we restrict $\phi_1$ accordingly, as in Equation~\eqref{eqn:restrict1}. Note that we do not replace atomic propositions protected by a $\X$ operator.
\begin{equation}
\neg ( \neg (\neg \top \wedge \X (q \U p)) \wedge \neg (\top \wedge \X (p \U (q \wedge \X (q \U p))))) \label{eqn:restrict1}
\end{equation}

After we minimize the resulting formula and advance one step, we obtain $\phi_2 = p \U (q \wedge \X (q \U p)))$. Since $\phi_2 = \phi_1$, we again calculate the immediate reward as zero.

Finally, after executing $a_2 =\text{back}$ randomly, we get out of the AUT, as shown in Figure~\ref{fig:episode1}, so the current activity is neither MainActivity nor AboutActivity. Therefore, the labeling is empty, so $p = q = \neg \top$. We calculate the final formula as $\phi_3 = \neg \top$. We calculate the immediate reward as $r = -1$ and terminate this episode.

In the second episode, FARLEAD-Android opens AboutActivity, but fails to return to MainActivity, as shown in Figure~\ref{fig:episode2}. Though it gets $r = -1$ in the end, it also obtains $r = .33$ for opening AboutActivity. The intermediate reward instructs FARLEAD-Android to explore the second action again in the next episode.

In the final episode, FARLEAD-Android again opens AboutActivity and gets $r = .33$ as before. This time, it finds the correct action and returns to MainActivity, as shown in Figure~\ref{fig:episode3}. Therefore, it receives $r = 1$ and terminates. The action sequence generated in the final episode is the satisfying test for the specification $\phi_0$.

\begin{table}
\caption{FARLEAD-Android Example}\label{tbl:example}
\ybox{.92}{!}{
\begin{tabular}{|r|l|l|l|l|}
\hline
\multicolumn{5}{l}{$\AP = \lbrace p = [\text{ activity}\sim\text{Main }], q = [\text{ activity}\sim\text{About }]\rbrace$} \\
\hline
\hline
& \multicolumn{2}{l|}{EPISODE $i = 1$} & \multicolumn{2}{l|}{$\phi_0 = \X (p \U (q \wedge \X (q \U p)))$} \\
\hline
$k = 0$ & $a_0 = $ reinit chesswalk MainActivity & $L = \lbrace p \rbrace$ & $\phi_1 = p \U (q \wedge \X (q \U p))$ & $r = 0$ \\
$k = 1$ & $a_1 = $ pauseresume & $L = \lbrace p \rbrace$ & $\phi_2 = p \U (q \wedge \X (q \U p))$ & $r = 0$ \\
$k = 2$ & $a_2 = $ back & $L = \lbrace \rbrace$ & $\phi_3 = \neg \top$ & $r = -1$ \\
\hline
& \multicolumn{2}{l|}{EPISODE $i = 2$} & \multicolumn{2}{l|}{$\phi_0 = \X (p \U (q \wedge \X (q \U p)))$} \\
\hline
$k = 0$ & $a_0 = $ reinit chesswalk MainActivity & $L = \lbrace p \rbrace$ & $\phi_1 = p \U (q \wedge \X (q \U p))$ & $r = 0$ \\
$k = 1$ & $a_1 = $ click 239 669 & $L = \lbrace q \rbrace$ & $\phi_2 = q \U p$ & $r = .33$ \\
$k = 2$ & $a_2 = $ click 143 32 & $L = \lbrace \rbrace$ & $\phi_3 = \neg \top$ & $r = -1$ \\
\hline
& \multicolumn{2}{l|}{EPISODE $i = 3$} & \multicolumn{2}{l|}{$\phi_0 = \X (p \U (q \wedge \X (q \U p)))$} \\
\hline
$k = 0$ & $a_0 = $ reinit chesswalk MainActivity & $L = \lbrace p \rbrace$ & $\phi_1 = p \U (q \wedge \X (q \U p))$ & $r = 0$ \\
$k = 1$ & $a_1 = $ click 239 669 & $L = \lbrace q \rbrace$ & $\phi_2 = q \U p$ & $r = .33$ \\
$k = 2$ & $a_2 = $ back & $L = \lbrace p \rbrace$ & $\phi_3 = \top$ & $r = 1$ \\
\hline
\end{tabular}}
\end{table}

\section{Evaluation}
\label{sec:eval}
In this section, we demonstrate the effectiveness and the performance of FARLEAD-Android through experiments on a VirtualBox guest with Android 4.4 operating system and 480x800 screen resolution. In evaluation, a virtual machine is better than a physical Android device because \begin{enumerate*} \item anyone can reproduce our experiments without the physical device, and \item even if a physical device is available, it must be the same with the original to produce the same results.\end{enumerate*} 

We downloaded two applications from F-Droid, namely ChessWalk and Notes. F-Droid \cite{Gultnieks:2010:FDROID} is an Android GUI application database, and many Android testing studies use it. We find F-Droid useful because it provides old versions and bug reports of the applications.

\newlength{\bls}
\setlength{\bls}{\baselineskip-\doublerulesep}
\begin{table*}
\caption{Our Experimental Specifications}\label{tbl:guifun}
\ybox{.958}{!}{
\begin{tabular}{|c|>{\centering\arraybackslash}m{0.5cm}|>{\raggedleft\arraybackslash}m{1.2cm}|m{4.5cm}|c|m{18cm}|}
\cline{3-6}
 \multicolumn{2}{c|}{} & \multicolumn{1}{c|}{\textbf{Source}} & \multicolumn{1}{c|}{\textbf{Description}} & \multicolumn{1}{c|}{\textbf{Level}} & \multicolumn{1}{c|}{\textbf{Formula}} \\
\hline
\parbox[t]{2mm}{\multirow{21}{*}{\rotatebox[origin=c]{90}{\textbf{AUT:} ChessWalk \hspace*{96pt}}}} & \multirow{3}{0.5cm}{\centering\arraybackslash $\phi_A$} & \multirow{3}{1cm}{\raggedleft\arraybackslash App-Agnostic} & \multirow{3}{4.5cm}{The user must be able to go to AboutActivity and return back.} & \centering\arraybackslash (a) & $\X ([\text{activity}\sim\text{Main}] \U ([\text{activity}\sim\text{About}] \wedge \X ([\text{activity}\sim\text{About}] \U [\text{activity}\sim\text{Main}]))$ \\ 
\cline{5-6}
& & & & (b) & $\X (([\text{activity}\sim\text{Main}] \wedge \text{actionType}=\text{click}) \U ([\text{activity}\sim\text{About}] \wedge \X ([\text{actionType}=\text{back}] \U [\text{activity}\sim\text{Main}]))$ \\ 
\cline{5-6}
& & & & (c) & $\X ((([\text{actionType}=\text{click}] \wedge [\text{actionDetail}\sim\text{About}]) \wedge [\text{activity}\sim\text{About}]) \wedge \X([\text{actionType}=\text{back}] \U \text{activity}\sim\text{Main}))$ \\
\cline{2-6}
& \multirow{3}{0.5cm}{\centering\arraybackslash $\phi_B$} & \multirow{3}{1cm}{\raggedleft\arraybackslash App-Agnostic} & \multirow{3}{4.5cm}{The user must be able to go to SettingsActivity and return back.} & \centering\arraybackslash (a) & $\X ([\text{activity}\sim\text{Main}] \U ([\text{activity}\sim\text{Settings}] \wedge \X ([\text{activity}\sim\text{Settings}] \U [\text{activity}\sim\text{Main}]))$ \\ 
\cline{5-6}
& & & & (b) & $\X (([\text{activity}\sim\text{Main}] \wedge \text{actionType}=\text{click}) \U ([\text{activity}\sim\text{Settings}] \wedge \X ([\text{actionType}=\text{back}] \U [\text{activity}\sim\text{Main}]))$ \\ 
\cline{5-6}
& & & & (c) & $\X ((([\text{actionType}=\text{click}] \wedge [\text{actionDetail}\sim\text{Settings}]) \wedge [\text{activity}\sim\text{Settings}]) \wedge \X([\text{actionType}=\text{back}] \U [\text{activity}\sim\text{Main}]))$ \\
\cline{2-6}
& \multirow{3}{0.5cm}{\centering\arraybackslash $\phi_C$} & \multirow{3}{1cm}{\raggedleft\arraybackslash App-Agnostic} & \multirow{3}{4.5cm}{Pausing and resuming the AUT should not change the screen.} & \centering\arraybackslash (a) & N/A \\ 
\cline{5-6}
& & & & (b) & $\X ([\text{actionType}=\text{pauseresume}] \wedge [\text{activity}\sim\text{Main}])$ \\ 
\cline{5-6}
& & & & (c) & $\X ([\text{actionType}=\text{pauseresume}] \wedge [\text{activity}\sim\text{Main}])$ \\ 
\cline{2-6}
& \multirow{3}{0.5cm}{\centering\arraybackslash $\phi_D$} & \multirow{3}{1cm}{\raggedleft\arraybackslash Bug Reports} & \multirow{3}{4.5cm}{The AUT should prevent the device from sleeping but it does not.} & \centering\arraybackslash (a) & $\top \U [\text{screen}=\text{off}]$ \\ 
\cline{5-6}
& & & & (b) & $\X ([\text{actionType}=\text{idle}] \wedge [\text{screen}=\text{off}])$ \\ 
\cline{5-6}
& & & & (c) & $\X ([\text{actionType}=\text{idle}] \wedge [\text{screen}=\text{off}])$ \\ 
\cline{2-6}
& \multirow{3}{0.5cm}{\centering\arraybackslash \vspace{6pt} \newline\newline $\phi_E$} & \multirow{3}{1cm}{\raggedleft\arraybackslash \vspace{-3pt} \newline \newline App-Agnostic} & \multirow{3}{4.5cm}{\vspace{-6pt} \newline \newline When the user changes some settings, the AUT should remember it later.} & \centering\arraybackslash (a) & $\X([\text{activity}\sim\text{Main}] \U (([\text{activity}\sim\text{Settings}] \wedge [\text{checked}=\text{true}]) \wedge (([\text{checked}=\text{true}] \wedge [\text{activity}\sim\text{Settings}]) \U ([\text{checked}=\text{false}] \wedge \X([\text{activity}\sim\text{Main}] \wedge \X ([\text{activity}\sim\text{Settings}] \wedge [\text{checked}=\text{false}]))))))$ \\ 
\cline{5-6}
& & & & (b) & $\X(([\text{actionType}=\text{click}] \wedge [\text{activity}\sim\text{Main}]) \U (([\text{activity}\sim\text{Settings}] \wedge [\text{checked}=\text{true}]) \wedge \X((([\text{actionType}=\text{click}] \wedge [\text{activity}\sim\text{Settings}]) \wedge [\text{checked}=\text{true}]) \U ([\text{checked}=\text{false}] \wedge \X(([\text{actionType}=\text{back}] \wedge [\text{activity}\sim\text{Main}]) \wedge \X(([\text{actionType}=\text{click}] \wedge [\text{activity}\sim\text{Main}]) \U ([\text{activity}\sim\text{Settings}] \wedge [\text{checked}=\text{false}])))))))$ \\ 
\cline{5-6}
& & & & (c) & $\X(([\text{actionDetail}\sim\text{Settings}] \wedge [\text{checked}=\text{true}]) \wedge \X((([\text{actionType}=\text{click}] \wedge [\text{actionObjectID}=\text{0:0:0:0:0}]) \wedge [\text{checked}=\text{false}]) \wedge \X([\text{actionType}=\text{back}] \wedge \X([\text{actionDetail}\sim\text{Settings}] \wedge [\text{checked}=\text{false}]))))$ \\ 
\cline{2-6}
& \multirow{3}{0.5cm}{\centering\arraybackslash \newline $\phi_F$} & \multirow{3}{1.2cm}{\raggedleft\arraybackslash \newline Manually Created} & \multirow{3}{4.5cm}{\newline The user must be able to start a game and make a move.} & \centering\arraybackslash (a) & $\X([\text{activity}\sim\text{New}] \wedge \X([\text{activity}\sim\text{Offline}] \wedge \X[\text{text}\sim\text{moved}]))$ \\ 
\cline{5-6}
& & & & (b) & $\X(([\text{actionType}=\text{click}] \wedge [\text{activity}\sim\text{New}]) \wedge \X(([\text{actionType}=\text{click}] \wedge [\text{activity}\sim\text{Offline}]) \wedge \X([\text{actionType}=\text{chessmove}] \wedge [\text{text}\sim\text{moved}])))$ \\ 
\cline{5-6}
& & & & (c) & $\X([\text{actionDetail}\sim\text{Offline}] \wedge \X([\text{actionDetail}\sim\text{Play}] \wedge \X([\text{actionType}=\text{chessmove}] \wedge [\text{text}\sim\text{moved}])))$ \\ 
\cline{2-6}
& \multirow{3}{0.5cm}{\centering\arraybackslash \vspace*{6pt}\newline \newline $\phi_G$} & \multirow{3}{1cm}{\raggedleft\arraybackslash \newline\newline Novel Bug} & \multirow{3}{4.5cm}{\vspace*{6pt}\newline When the user starts a second game, moves of the first game is shown.} & \centering\arraybackslash (a) & $\X([\text{activity}\sim\text{Main}] \U ([\text{activity}\sim\text{New}] \wedge \X([\text{activity}\sim\text{New}] \U ([\text{activity}\sim\text{Offline}] \wedge \X([\text{activity}\sim\text{Offline}] \U ([\text{text}\sim\text{moved}] \wedge \X([\text{activity}\sim\text{Offline}] \U ([\text{activity}\sim\text{New}] \wedge \X([\text{activity}\sim\text{New}] \U ([\text{activity}\sim\text{Offline}] \wedge [\text{text}\sim\text{moved}]))))))))))$ \\ 
\cline{5-6}
& & & & (b) & $\X(([\text{activity}\sim\text{Main}] \wedge [\text{actionType}=\text{click}]) U ([\text{activity}\sim\text{New}] \wedge \X(([\text{actionType}=\text{click}] \wedge [\text{activity}\sim\text{Offline}]) \wedge \X(([\text{actionType}=\text{chessmove}] \wedge [\text{text}\sim\text{moved}]) \wedge \X([\text{actionType}=\text{click}] U ([\text{activity}\sim\text{New}] \wedge \X(([\text{actionType}=\text{click}] \wedge [\text{activity}\sim\text{Offline}]) \wedge [\text{text}\sim\text{moved}])))))))$ \\ 
\cline{5-6}
& & & & (c) & $\X([\text{actionDetail}\sim\text{Offline}] \wedge \X([\text{actionDetail}\sim\text{Play}] \wedge \X([\text{actionType}=\text{chessmove}] \wedge \X([\text{actionDetail}\sim\text{New}] \wedge \X([\text{actionDetail}\sim\text{Play}] \wedge [\text{text}\sim\text{moved}])))))$ \\ 
\cline{1-6}\multicolumn{2}{c}{}&\\[-\bls]\cline{3-6}
 \multicolumn{2}{c|}{} & \multicolumn{1}{c|}{\textbf{Source}} & \multicolumn{1}{c|}{\textbf{Description}} & \multicolumn{1}{c|}{\textbf{Level}} & \multicolumn{1}{c|}{\textbf{Formula}} \\
\hline
\parbox[t]{2mm}{\multirow{6}{*}{\rotatebox[origin=c]{90}{\textbf{AUT:} Notes \hspace*{36pt}}}} & \multirow{3}{0.5cm}{\centering\arraybackslash \vspace*{6pt}\newline \newline $\phi_H$} & \multirow{3}{1cm}{\raggedleft\arraybackslash \newline\newline Novel Bug} & \multirow{3}{4.5cm}{\vspace*{6pt}\newline The sketch note palette does not show one of the 20 color choices (black is missing).} & \centering\arraybackslash (a) & $\X([\text{text}\sim\text{OK}] \U ([\text{activity}\sim\text{Main}] \wedge \X([\text{activity}\sim\text{Main}] \U ([\text{text}\sim\text{sketch}] \wedge \X([\text{activity}\sim\text{Main}] \U ([\text{activity}\sim\text{Sketch}] \wedge \X([\text{activity}\sim\text{Sketch}] \U ([\text{objectID}\sim\text{18}] \wedge \neg[\text{objectID}\sim\text{19}]))))))))$ \\ 
\cline{5-6}
& & & & (b) & $\X(([\text{text}\sim\text{OK}] \wedge [\text{actionType}=\text{click}]) \U ([\text{activity}\sim\text{Main}] \wedge \X(([\text{activity}\sim\text{Main}] \wedge [\text{actionType}=\text{click}]) \U ([\text{text}\sim\text{sketch}] \wedge \X(([\text{activity}\sim\text{Main}] \wedge [\text{actionType}=\text{click}]) \U ([\text{activity}\sim\text{Sketch}] \wedge \X(([\text{activity}\sim\text{Sketch}] \wedge [\text{actionType}=\text{click}]) \U ([\text{objectID}\sim\text{18}] \wedge \neg[\text{objectID}\sim\text{19}]))))))))$ \\ 
\cline{5-6}
& & & & (c) & $\X(([\text{actionType}=\text{click}] \wedge [\text{actionDetail}\sim\text{OK}]) \wedge \X(([\text{actionType}=\text{click}] \wedge [\text{actionDetail}\sim\text{+}]) \wedge \X(([\text{actionType}=\text{click}] \wedge [\text{actionDetail}\sim\text{sketch}]) \wedge \X(([\text{actionType}=\text{click}] \wedge [\text{actionDetail}\sim\text{colorSelector}]) \U ([\text{objectID}\sim\text{18}] \wedge \neg [\text{objectID}\sim\text{19}])))))$ \\ 
\cline{2-6}
& \multirow{3}{0.5cm}{\centering\arraybackslash \vspace*{6pt}\newline $\phi_I$} & \multirow{3}{1cm}{\raggedleft\arraybackslash \vspace*{3pt}\newline Bug Reports} & \multirow{3}{4.5cm}{ \vspace*{3pt}\newline Even if the user cancels a note, a dummy note is still created.} & \centering\arraybackslash (a) & N/A \\ 
\cline{5-6}
& & & & (b) & $\X(([\text{activity}\sim\text{Main}] \wedge [\text{actionType}=\text{click}]) \U ([\text{text}\sim\text{New text}] \wedge \X(([\text{actionType}=\text{click}] \wedge [\text{activity}\sim\text{TextNote}]) \wedge \X([\text{actionType}=\text{back}] \U [\text{text}\sim\text{Note 1}]))))$ \\ 
\cline{5-6}
& & & & (c) & $\X(([\text{actionType}=\text{click}] \wedge [\text{actionDetail}\sim\text{OK}]) \wedge \X(([\text{actionType}=\text{click}] \wedge [\text{actionDetail}\sim\text{+}]) \wedge \X(([\text{actionType}=\text{click}] \wedge [\text{actionDetail}\sim\text{text}]) \wedge \X([\text{actionType}=\text{back}] \U [\text{text}\sim\text{Note 1}]))))$ \\ 
\hline
\end{tabular}
}
\end{table*}

Table~\ref{tbl:guifun} shows the GUI-level specifications we obtained for ChessWalk and Notes applications. These specifications come from four sources, \begin{enumerate*} \item app-agnostic test oracles \cite{Zaeem+:2014:ICST}, \item bug reports in the F-Droid database, \item novel bugs we found, and \item specifications we manually created.\end{enumerate*} Note that we can specify the same GUI function with different LTL formulae. In an LTL formula, an action label starts with the word $action$. Otherwise, it is a state label. There are two kinds of action labels, type and detail. An action type label constrains the action type, while an action detail label constrains the action parameters. Using these label categories, we define three levels of detail for LTL formulae with (a) only state labels, (b) state labels and action type labels, and (c) all labels. Intuitively, FARLEAD-Android should be more effective as the level of detail goes from (a) to (c). Note that specifications $\phi_C$ and $\phi_I$ are inexpressable with a level (a) formula because they explicitly depend on action labels.

We investigate FARLEAD-Android in three categories, FARLEADa, FARLEADb, and FARLEADc, indicating that we use a level (a), (b), or (c) formula, respectively. For specifications $\phi_C$ and $\phi_D$, the LTL formula does not change from level (b) to (c) because the specified action does not take any parameters. Hence, we combine FARLEADb and FARLEADc as FARLEADb/c for these specifications. Other than FARLEAD-Android, we perform experiments on three known approaches, \begin{enumerate*} \item random exploration (Random), \item Google's built-in monkey tester (Monkey) \cite{Google:MONKEY}, and \item Q-Learning Based Exploration optimized for activity coverage (QBEa) \cite{Koroglu+:2018:ICST}.\end{enumerate*} Random explores the AUT with completely random actions using the same action set of FARLEAD-Android. Monkey also explores the AUT randomly, but with its own action set. QBEa chooses actions according to a pre-learned probability distribution optimized for traversing activities. We implement these approaches in FARLEAD-Android so we can check if they satisfy our specifications, on-the-fly.

For every specification in Table~\ref{tbl:guifun}, we execute Random, Monkey, QBEa, FARLEADa, FARLEADb, and FARLEADc $100$ times each for a maximum of $E = 500$ episodes. The maximum number of steps is $K = 4$ or $K = 6$, depending on the specification, so every execution runs up to $500$ episodes with at most six steps per episode. We keep the remaining parameters of FARLEAD-Android fixed throughout our experiments. We use these experiments to evaluate the effectiveness and the performance of FARLEAD-Android in Sections~\ref{sec:eval:subsec:effectiveness} and \ref{sec:eval:subsec:performance}, respectively. In Section~\ref{sec:eval:subsec:levels}, we discuss the impact of the detail levels on the effectiveness/performance. In Section~\ref{sec:eval:subsec:steps}, we compare the number of steps taken in our experiments with known RL-LTL methods.

\begin{table}
\caption{Effectiveness of Engines per Specification}\label{tbl:effectiveness}
\ybox{.8}{!}{\begin{tabular}{|r|ccccccccc|c|}
\hline
\textbf{Engine} & $\phi_A$ & $\phi_B$ & $\phi_C$ & $\phi_D$ & $\phi_E$ & $\phi_F$ & $\phi_G$ & $\phi_H$ & $\phi_I$ & \textbf{Total} \\
\hline
Random & \checkmark & \checkmark & \checkmark & \checkmark & \checkmark & \checkmark &  &  & & 6 \\
\hline
Monkey & \checkmark & \checkmark & & & & & \checkmark & & & 3 \\
\hline
QBEa & \checkmark & \checkmark & & & \checkmark & & & \checkmark & & 4 \\
\hline
FARLEADa & \checkmark & \checkmark & & \checkmark & \checkmark & \checkmark & \checkmark & \checkmark & & 7 \\
\hline
FARLEADb & \checkmark & \checkmark & \checkmark & \checkmark & \checkmark & \checkmark & \checkmark & \checkmark & \checkmark & 9 \\
\hline
FARLEADc & \checkmark & \checkmark & \checkmark & \checkmark & \checkmark & \checkmark & \checkmark & \checkmark & \checkmark & 9 \\
\hline
\end{tabular}}
\end{table}

\subsection{Effectiveness}
\label{sec:eval:subsec:effectiveness}

We say that an engine was effective at satisfying a GUI-level specification if it generated a satisfying test at least once in our experiments.

Table~\ref{tbl:effectiveness} shows the total number of specifications that our engines were effective at satisfying. Our results show that FARLEADa, FARLEADb, and FARLEADc were effective at more specifications than Random, Monkey, and QBEa. Hence, we conclude that FARLEAD-Android is more effective than other engines. 

\begin{figure*}
\includegraphics[width=.95\linewidth]{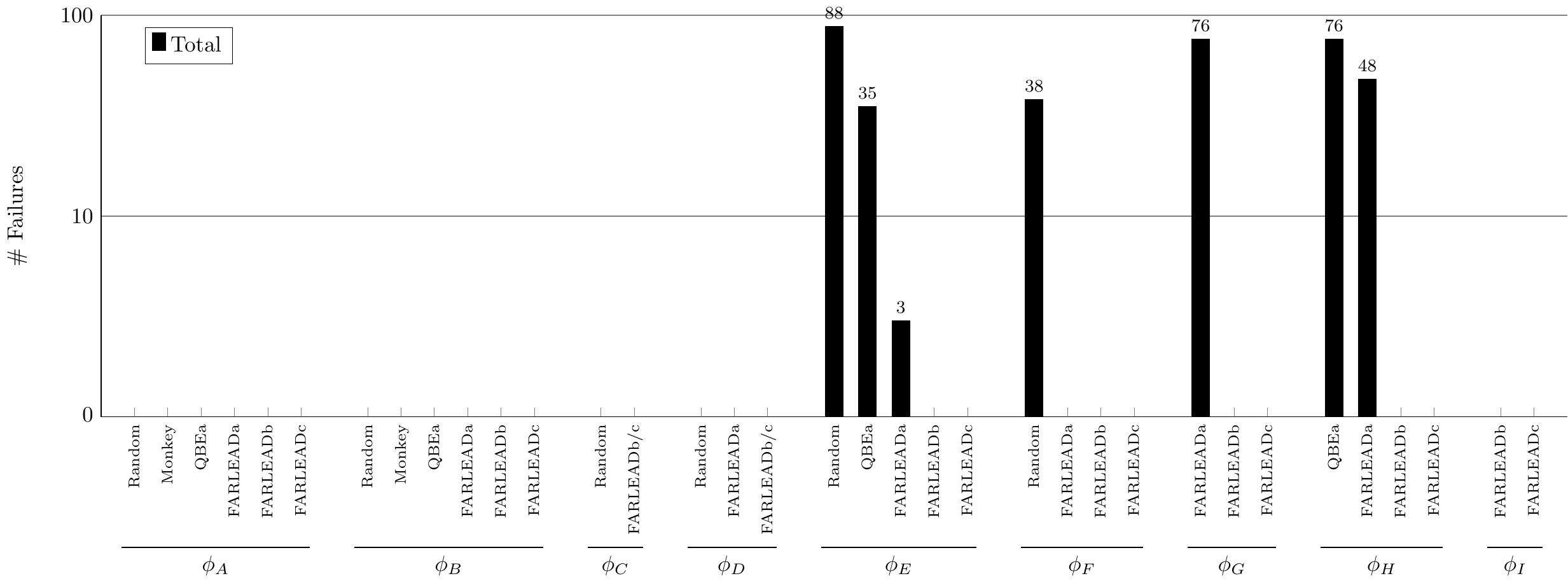}
\caption{Number of Failures Across 100 Executions}\label{fig:fails}
\Description{Number of Failures Across 100 Executions}
\end{figure*}

\begin{figure*}
\includegraphics[width=.95\linewidth]{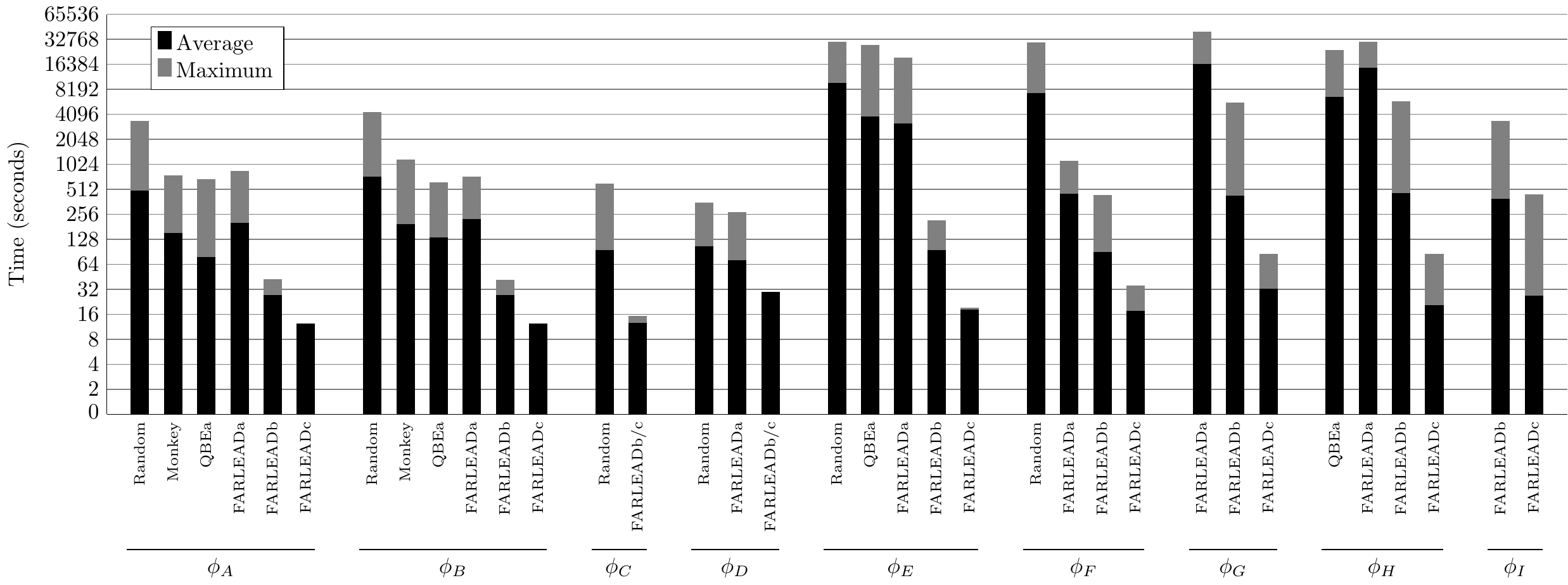}
\caption{Test Times Across 100 Executions}\label{fig:times}
\Description{Test Times Across 100 Executions}
\end{figure*}

\begin{figure*}
\includegraphics[width=.95\linewidth]{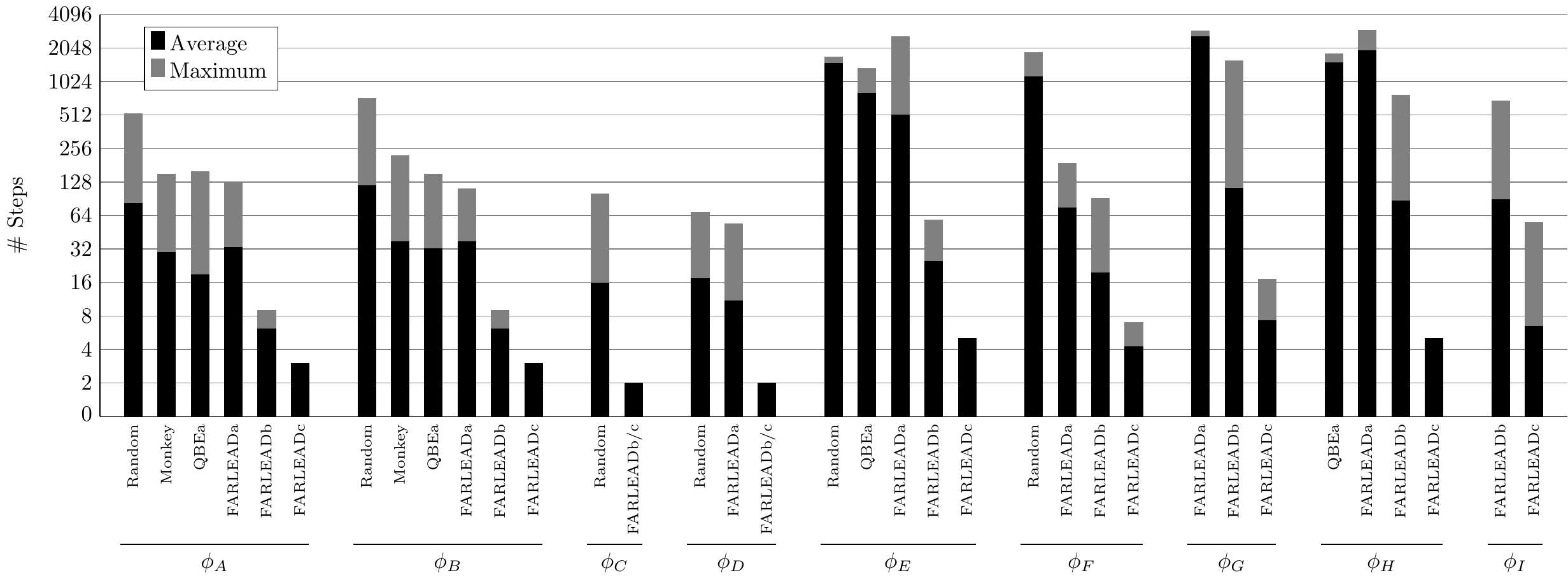}
\caption{Number of Steps Across 100 Executions}\label{fig:steps}
\Description{Number of Steps Across 100 Executions}
\end{figure*}

\subsection{Performance}
\label{sec:eval:subsec:performance}

We say that an engine failed to satisfy a GUI-level specification if it could not generate a satisfying test in one execution. We consider an engine achieved higher performance than another if it failed fewer times in our experiments. 

Figure~\ref{fig:fails} shows the number of failures of all the engines across $100$ executions in logarithmic scale. According to this figure, FARLEADa, FARLEADb, and FARLEADc failed fewer times than Random, Monkey, and QBEa at $\phi_E$, $\phi_F$, and $\phi_H$, indicating that FARLE\-AD-Android achieved higher performance for these specifications. Only FARLEAD-Android was effective at $\phi_G$ and $\phi_I$, so we ignore those specifications in evaluating performance.

Figure~\ref{fig:times} shows the average and the maximum times required to terminate for all the engines with every specification across $100$ executions in logarithmic scale. According to this figure, FARLEADb and FARLEADc spent less time on average and in the worst case than Random, Monkey, and QBEa for the remaining specifications $\phi_A$-$\phi_D$, indicating that FARLEAD-Android achieved higher performance when it used a level (b) or (c) formula. However, our results show that QBEa and Monkey spent less time than FARLEADa for $\phi_A$ and $\phi_B$ because these specifications are about traversing activities only, a task which QBEa explicitly specializes on and Monkey excels at [5]. As a last note, FARLEADa outperformed QBEa for $\phi_H$ even though QBEa spent less time than FARLEADa because QBEa failed more than FARLEADa. Hence, we conclude that FARLEAD-Android achieves higher performance than Random, Monkey, and QBEa unless the specification is at level (a) and about traversing activities only.

\subsection{Impact of LTL Formula Levels}
\label{sec:eval:subsec:levels}

Table~\ref{tbl:effectiveness} shows that FARLEADb and FARLEADc were effective at more specifications than FARLEADa. Hence, we conclude that FARLEAD-Android becomes more effective when the level of detail goes from (a) to (b) or (c).

Figure~\ref{fig:fails} shows that FARLEADa failed more times than FARLEADb and FARLEADc at $\phi_E$-$\phi_H$, indicating that FARLEADb and FARLEADc outperformed FARLEADa. FARLEADa was not effective at $\phi_C$ and $\phi_I$, so we ignore those specifications in evaluating performance. Figure~\ref{fig:times} shows that FARLEADa spent more time than FARLEADb and FARLEADc at the remaining specifications, $\phi_A$, $\phi_B$, and $\phi_D$, indicating that FARLEADb and FARLEADc again outperformed FARLEADa. Furthermore, Figure~\ref{fig:times} shows that FARLEADc spent less time than FARLEADb in all specifications except $\phi_C$ and $\phi_D$, where FARLEADb and FARLEADc are equivalent. Hence, we conclude that the performance of FARLEAD-Android increases as the level of detail goes from (a) to (c).

Overall, both the effectiveness and the performance of FARLEAD-Android increase with the level of detail in LTL formula. Hence, we recommend developers to provide as much detail as possible to get the highest effectiveness and performance from FARLEAD-Android.

\subsection{Number of Steps to Terminate}
\label{sec:eval:subsec:steps}

Figure~\ref{fig:steps} shows the average and the maximum number of steps required to terminate for all the engines with every specification across $100$ executions in logarithmic scale. The number of steps is a known measure used to compare RL methods logically constrained with LTL formulae \cite{Hasanbeig+:2019:AAMAS,Hasanbeig+:2019:arXiv,Icarte+:2018:AAMAS,Wen+:2015:IROS}. Known RL-LTL methods take a high number of steps, in the order of hundreds of thousands, because these methods aim to converge to an optimal policy. FARLEAD-Android took less than four thousand steps in all experiments since it terminates as soon as it finds a satisfying test, which occurs before convergence. 
Figures~\ref{fig:times} and \ref{fig:steps} show that the number of steps is a similar metric to the test times, indicating that the number of steps is suitable for evaluating performance. We evaluate performance using the test times instead of the number of steps because the test times will directly affect the developer's ability to test more.

\section{Related Work}
\label{sec:relwork}
Several studies \cite{Hasanbeig+:2019:AAMAS,Hasanbeig+:2019:arXiv,Icarte+:2018:AAMAS,Wen+:2015:IROS} use an LTL specification as a high-level guide for an RL agent. The RL agent in these studies never terminate and has to avoid violating a given specification indefinitely. For this purpose, they define the LTL semantics over infinite traces and train the RL agent until it converges to an optimal policy. Convergence ensures that the RL agent will continue to satisfy the specification. The goal of FARLEAD-Android is not to train an RL agent until it converges to an optimal policy but to generate one satisfying test and terminate before the convergence occurs. Hence, we define the LTL semantics over finite traces. As a result, FARLEAD-Android requires much fewer steps, which is crucial due to the high execution cost of Android devices. We also develop many improvements to the RL algorithm, so FARLEAD-Android generates a satisfying test with as few steps as possible. To the best of our knowledge, FARLEAD-Android is the first engine that combines RL and LTL for test generation.

Two studies \cite{Behjati+:2009:FSE,AraragiCho:2006:Springer} use RL for model checking LTL properties. These studies do not prove LTL properties but focus on finding counterexamples. This type of model checking can replace testing given an appropriate model. However, such models are not readily available for Android applications, and it is hard to generate them.

Two studies \cite{Koroglu+:2018:ICST,Mariani+:2012:ICST} use RL to generate tests for GUI applications. Although engines proposed in these studies are automated, they do not check if they test a target GUI function or not. In literature, this is a known issue called the oracle problem. We propose a solution to this problem by specifying GUI functions as monitorable LTL formulae.

Monitoring LTL properties dynamically from an Android device is not a new idea. To the best of our knowledge, there are three Android monitoring tools, RV-Droid \cite{Falcone+:2012:RV}, RV-Android \cite{Daian+:2015:RV}, and ADRENALIN-RV \cite{Sun+:2017:ICST}. All these tools monitor LTL properties at the source code level. Instead, our monitoring is at the GUI level. Furthermore, these tools do not generate tests. They assume that a test is given and focus only on monitoring properties. Instead, FARLEAD-Android performs  test generation and monitoring at the same time.

FARLEAD-Android monitors an LTL specification by making changes to it at every step using the \emph{projection} procedure. This kind of LTL monitoring is called LTL progression (formula rewriting). Icarte et al. \cite{Icarte+:2018:AAMAS} and RV-Droid \cite{Falcone+:2012:RV} use LTL progression. As an alternative, RV-Android \cite{Daian+:2015:RV} first translates an LTL specification into a Past Time Linear Temporal Logic (PTLTL) formula and monitors past actions using that formula. Using LTL progression enables us to use Reward Shaping \cite{Laud:2004}. To the best of our knowledge, FARLEAD-Android is the first engine that uses Reward Shaping in RL-LTL.

\section{Discussion}
\label{sec:discussion}
Effectiveness of FARLEAD-Android depends heavily on the exact definitions of atomic propositions, GUI states, and actions. We believe FARLEAD-Android will be able to generate tests for more GUI functions as these definitions get richer. To the best of our abilities, we made these definitions as comprehensive as possible. 

Android devices and applications are non-deterministic. As a result, the same tes t sometimes gets different rewards. RL is known to be robust against non-determinism, so this was not a problem in our experiments. However, due to the non-determinism, a satisfying test $ts$ may not always generate the same execution trace $t$ and therefore may not always satisfy a given LTL formula $\phi$. We say that a satisfying test $ts$ is reliable under an LTL formula $\phi$ if and only if all possible execution traces of the test $ts$ satisfy the LTL formula $\phi$. Replaying the test $ts$ several times may help to establish confidence in test reliability.

FARLEAD-Android is trivially sound because it guarantees that the specification is satisfied if termination occurs before the maximum number of episodes.
However, it is not complete because it cannot decide unsatisfiability. For this purpose, if FARLEAD-Android terminates without producing a satisfying test, we could either use a model checker or warn the developer to investigate the associated GUI function manually. 

FARLEAD-Android is fully automated, does not need the source code of the AUT, and does not instrument the AUT. FARLEAD-Android requires the developer to have experience in LTL. A converter that transforms a user-friendly language, for example, Gherk\-in Syntax, to LTL could make FARLEAD-Android even more practical for the developer.

When the LTL formula is of level (c), it usually takes a single episode to generate a satisfying test. It is possible to use this fact to store tests as level (c) LTL formulae instead of action sequences. Storing tests as LTL has two advantages. First, it is more portable where GUI actions may be too specific. For example, a \emph{click} requires two coordinates that could be different for various devices. However, one can abstract these details in a single LTL formula and obtain tests for as many devices as possible. Second, the LTL formula encodes the test oracle in addition to the test, whereas the test only specifies the action sequence and not what should we check at the end. We believe these advantages are side benefits of our approach.

Although an RL agent typically stores the quality function $Q$ as a look-up table, in theory, any regression method (function approximator) can replace the look-up table \cite[p.533]{Alpaydin:2014:ML}. If the function approximator is an artificial neural network, then this approach is called Deep Reinforcement Learning (Deep RL) \cite{Mnih+:2013:arXiv}. Deep Double SARSA and Deep Double Expected SARSA are two example variants of the original Deep RL \cite{Ganger+:2016:JDAIP}. Deep RL requires large amounts of data to train and therefore, may result  in high execution cost.

We could add fatal exceptions to FARLEAD-Android as a contextual attribute of the current state, such as $\text{crashed}=\text{true}|\text{false}$. So, $\phi_{\text{crash}} = \top \U [\text{crashed}=\text{true}]$ will effectively look for a crash. However, this is not the intended use of FARLEAD-Android. With $\phi_{\text{crash}}$, we expect  FARLEAD-Android to be less effective than a test generation engine optimized for crash detection. However, FARLEAD-Android can be effective at reproducing known bugs, as we show in our experiments. We have also found two new bugs while investigating our experimental AUTs. We have reported these bugs to the respective developers of these AUTs using the issue tracking system.

We can specify the reachability of every activity in the AUT using LTL. Then, the tests FARLEAD-Android generate will achieve $100$\% activity coverage, assuming all activities of the AUT are indeed reachable. However, this approach may fail to test many essential GUI functions of the AUT, as we state in Section~\ref{sec:intro}. 

Though we have experimented only on virtual machines, FARLE\-AD-Android also supports physical Android devices, the Android Emulator, and any abstract model that receives the actions and returns the observations described in Figure~\ref{fig:overview}. 

To evaluate a test generation tool in terms of its effectiveness at testing a GUI function, we must modify it for monitoring the LTL specification, on-the-fly. Such a modification requires a high amount of engineering, so we implemented the simple Random and Monkey approaches in FARLEAD-Android. We also included QBEa since FARLEAD-Android already had the support for RL, which QBEa uses. Since Monkey and QBEa are known to be efficient in achieving high activity coverage, we expect that comparisons with other test generation tools will also yield similar results.

Finally, many external factors may affect our experimental results. Examples are the Android OS version, issues with the Android Debugging Bridge (ADB), the size of the AUT, and the complexity of the specified GUI function. Therefore, we believe that FARLEAD-Android would benefit from replication studies under various conditions.

\section{Conclusion}
\label{sec:conclusion}
In this study, we proposed the Fully Automated Reinforcement LEArning-Driven Specification-Based Test Generator for Android (FARLEAD-Android). FARLEAD-Android uses Reinforcement Learning (RL) to generate replayable tests that satisfy given Linear-time Temporal Logic (LTL) specifications. To the best of our knowledge, FARLEAD-Android is the first test generation engine that combines RL and LTL. Our evaluation shows that FARLEAD-Android has been more effective and has achieved higher performance in generating tests that satisfy given specifications than three known test generation approaches, Random, Monkey, and QBEa. We also demonstrated that the developer should provide as much detail as possible to get the highest effectiveness and performance from FARLEAD-Android. Finally, we showed that the early termination of FARLEAD-Android allows us to take much fewer steps than known RL-LTL methods.

In the future, with FARLEAD-Android, we aim to generate tests for larger applications and more specifications. Currently, we cannot specify GUI functions that depend on timings. Hence, we aim to support Metric Temporal Logic (MTL), which will increase the number of GUI functions that we can specify. Finally, we will improve FARLEAD-Android so that it will support atomic propositions on sensory inputs, energy bugs, and security issues.\vspace*{6pt}

\bibliographystyle{ACM-Reference-Format}
\bibliography{input/citations2}


\begin{thebibliography}{49}


\ifx \showCODEN    \undefined \def \showCODEN     #1{\unskip}     \fi
\ifx \showDOI      \undefined \def \showDOI       #1{#1}\fi
\ifx \showISBNx    \undefined \def \showISBNx     #1{\unskip}     \fi
\ifx \showISBNxiii \undefined \def \showISBNxiii  #1{\unskip}     \fi
\ifx \showISSN     \undefined \def \showISSN      #1{\unskip}     \fi
\ifx \showLCCN     \undefined \def \showLCCN      #1{\unskip}     \fi
\ifx \shownote     \undefined \def \shownote      #1{#1}          \fi
\ifx \showarticletitle \undefined \def \showarticletitle #1{#1}   \fi
\ifx \showURL      \undefined \def \showURL       {\relax}        \fi
\providecommand\bibfield[2]{#2}
\providecommand\bibinfo[2]{#2}
\providecommand\natexlab[1]{#1}
\providecommand\showeprint[2][]{arXiv:#2}

\bibitem[\protect\citeauthoryear{Alpaydin}{Alpaydin}{2014}]%
        {Alpaydin:2014:ML}
\bibfield{author}{\bibinfo{person}{Ethem Alpaydin}.}
  \bibinfo{year}{2014}\natexlab{}.
\newblock \bibinfo{booktitle}{\emph{Introduction to Machine Learning}
  (\bibinfo{edition}{3rd} ed.)}.
\newblock \bibinfo{publisher}{The MIT Press}.
\newblock
\showISBNx{0262028182, 9780262028189}


\bibitem[\protect\citeauthoryear{Amalfitano, Fasolino, Tramontana, Ta, and
  Memon}{Amalfitano et~al\mbox{.}}{2015}]%
        {Amalfitano+:2015:IEEE}
\bibfield{author}{\bibinfo{person}{D. Amalfitano}, \bibinfo{person}{A.~R.
  Fasolino}, \bibinfo{person}{P. Tramontana}, \bibinfo{person}{B.~D. Ta}, {and}
  \bibinfo{person}{A.~M. Memon}.} \bibinfo{year}{2015}\natexlab{}.
\newblock \showarticletitle{{MobiGUITAR: Automated Model-Based Testing of
  Mobile Apps}}.
\newblock \bibinfo{journal}{\emph{IEEE Software}} \bibinfo{volume}{32},
  \bibinfo{number}{5} (\bibinfo{year}{2015}), \bibinfo{pages}{53--59}.
\newblock


\bibitem[\protect\citeauthoryear{Anand, Naik, Harrold, and Yang}{Anand
  et~al\mbox{.}}{2012}]%
        {Anand+:2012:FSE}
\bibfield{author}{\bibinfo{person}{Saswat Anand}, \bibinfo{person}{Mayur Naik},
  \bibinfo{person}{Mary~Jean Harrold}, {and} \bibinfo{person}{Hongseok Yang}.}
  \bibinfo{year}{2012}\natexlab{}.
\newblock \showarticletitle{Automated concolic testing of smartphone apps}. In
  \bibinfo{booktitle}{\emph{Proceedings of the ACM SIGSOFT 20th International
  Symposium on the Foundations of Software Engineering (FSE)}}.
\newblock
\newblock
\shownote{{https://github.com/saswatanand/acteve}.}


\bibitem[\protect\citeauthoryear{Araragi and Cho}{Araragi and Cho}{2006}]%
        {AraragiCho:2006:Springer}
\bibfield{author}{\bibinfo{person}{Tadashi Araragi} {and}
  \bibinfo{person}{Seung~Mo Cho}.} \bibinfo{year}{2006}\natexlab{}.
\newblock \showarticletitle{Checking liveness properties of concurrent systems
  by reinforcement learning}. In \bibinfo{booktitle}{\emph{International
  Workshop on Model Checking and Artificial Intelligence}}. Springer,
  \bibinfo{pages}{84--94}.
\newblock


\bibitem[\protect\citeauthoryear{Arel, Liu, Urbanik, and Kohls}{Arel
  et~al\mbox{.}}{2010}]%
        {Arel+:2010:IET}
\bibfield{author}{\bibinfo{person}{Itamar Arel}, \bibinfo{person}{Cong Liu},
  \bibinfo{person}{T Urbanik}, {and} \bibinfo{person}{AG Kohls}.}
  \bibinfo{year}{2010}\natexlab{}.
\newblock \showarticletitle{Reinforcement learning-based multi-agent system for
  network traffic signal control}.
\newblock \bibinfo{journal}{\emph{IET Intelligent Transport Systems}}
  \bibinfo{volume}{4}, \bibinfo{number}{2} (\bibinfo{year}{2010}),
  \bibinfo{pages}{128--135}.
\newblock


\bibitem[\protect\citeauthoryear{Azim and Neamtiu}{Azim and Neamtiu}{2013}]%
        {AzimNeamtiu:2013:OOPSLA}
\bibfield{author}{\bibinfo{person}{Tanzirul Azim} {and} \bibinfo{person}{Iulian
  Neamtiu}.} \bibinfo{year}{2013}\natexlab{}.
\newblock \showarticletitle{{Targeted and Depth-first Exploration for
  Systematic Testing of Android Apps}}. In \bibinfo{booktitle}{\emph{ACM
  SIGPLAN International Conference on Object Oriented Programming Systems
  Languages and Applications (OOPSLA)}}. \bibinfo{pages}{641--660}.
\newblock


\bibitem[\protect\citeauthoryear{Barr, Harman, McMinn, Shahbaz, and Yoo}{Barr
  et~al\mbox{.}}{2014}]%
        {Barr+:2014:TSE}
\bibfield{author}{\bibinfo{person}{Earl~T Barr}, \bibinfo{person}{Mark Harman},
  \bibinfo{person}{Phil McMinn}, \bibinfo{person}{Muzammil Shahbaz}, {and}
  \bibinfo{person}{Shin Yoo}.} \bibinfo{year}{2014}\natexlab{}.
\newblock \showarticletitle{The oracle problem in software testing: A survey}.
\newblock \bibinfo{journal}{\emph{IEEE transactions on software engineering}}
  \bibinfo{volume}{41}, \bibinfo{number}{5} (\bibinfo{year}{2014}),
  \bibinfo{pages}{507--525}.
\newblock


\bibitem[\protect\citeauthoryear{Behjati, Sirjani, and Ahmadabadi}{Behjati
  et~al\mbox{.}}{2009}]%
        {Behjati+:2009:FSE}
\bibfield{author}{\bibinfo{person}{Razieh Behjati}, \bibinfo{person}{Marjan
  Sirjani}, {and} \bibinfo{person}{Majid~Nili Ahmadabadi}.}
  \bibinfo{year}{2009}\natexlab{}.
\newblock \showarticletitle{Bounded rational search for on-the-fly model
  checking of LTL properties}. In \bibinfo{booktitle}{\emph{International
  Conference on Fundamentals of Software Engineering}}. Springer,
  \bibinfo{pages}{292--307}.
\newblock


\bibitem[\protect\citeauthoryear{Bolton}{Bolton}{2017}]%
        {APPLAUSE}
\bibfield{author}{\bibinfo{person}{David Bolton}.}
  \bibinfo{year}{2017}\natexlab{}.
\newblock \bibinfo{title}{88 percent of people will abandon an app because of
  bugs}.
\newblock
\newblock
\newblock
\shownote{https://www.applause.com/blog/app-abandonment-bug-testing.}


\bibitem[\protect\citeauthoryear{{Cao}, {Deng}, {Yu}, {Duan}, and {Ma}}{{Cao}
  et~al\mbox{.}}{2019}]%
        {Cao+:2019:COMPSAC}
\bibfield{author}{\bibinfo{person}{C. {Cao}}, \bibinfo{person}{J. {Deng}},
  \bibinfo{person}{P. {Yu}}, \bibinfo{person}{Z. {Duan}}, {and}
  \bibinfo{person}{X. {Ma}}.} \bibinfo{year}{2019}\natexlab{}.
\newblock \showarticletitle{ParaAim: Testing Android Applications Parallel at
  Activity Granularity}. In \bibinfo{booktitle}{\emph{2019 IEEE 43rd Annual
  Computer Software and Applications Conference (COMPSAC)}}.
\newblock


\bibitem[\protect\citeauthoryear{Cao, Wu, Chen, and Wei}{Cao
  et~al\mbox{.}}{2018}]%
        {Cao+:2018:Internetware}
\bibfield{author}{\bibinfo{person}{Yuzhong Cao}, \bibinfo{person}{Guoquan Wu},
  \bibinfo{person}{Wei Chen}, {and} \bibinfo{person}{Jun Wei}.}
  \bibinfo{year}{2018}\natexlab{}.
\newblock \showarticletitle{CrawlDroid: Effective Model-based GUI Testing of
  Android Apps}. In \bibinfo{booktitle}{\emph{Tenth Asia-Pacific Symposium on
  Internetware}}.
\newblock
\newblock
\shownote{{https://github.com/sy1121/CrawlDroid}.}


\bibitem[\protect\citeauthoryear{Choi}{Choi}{[n.d.]}]%
        {Choi:SWIFTHAND2}
\bibfield{author}{\bibinfo{person}{Wontae Choi}.}
  \bibinfo{year}{[n.d.]}\natexlab{}.
\newblock \bibinfo{title}{{SwiftHand2: Android GUI Testing Framework}}.
\newblock
\newblock
\newblock
\shownote{{\\https://github.com/wtchoi/swifthand2}.}


\bibitem[\protect\citeauthoryear{Choi, Necula, and Sen}{Choi
  et~al\mbox{.}}{2013}]%
        {Choi+:2013:OOPSLA}
\bibfield{author}{\bibinfo{person}{Wontae Choi}, \bibinfo{person}{George
  Necula}, {and} \bibinfo{person}{Koushik Sen}.}
  \bibinfo{year}{2013}\natexlab{}.
\newblock \showarticletitle{{Guided GUI Testing of Android Apps with Minimal
  Restart and Approximate Learning}}. In \bibinfo{booktitle}{\emph{ACM SIGPLAN
  International Conference on Object Oriented Programming Systems Languages and
  Applications (OOPSLA)}}. \bibinfo{pages}{623--640}.
\newblock


\bibitem[\protect\citeauthoryear{Daian, Falcone, Meredith, Serbanuta,
  Shiraishi, Iwai, and Rosu}{Daian et~al\mbox{.}}{2015}]%
        {Daian+:2015:RV}
\bibfield{author}{\bibinfo{person}{Philip Daian}, \bibinfo{person}{Yli{\`{e}}s
  Falcone}, \bibinfo{person}{Patrick~O'Neil Meredith},
  \bibinfo{person}{Traian{-}Florin Serbanuta}, \bibinfo{person}{Shinichi
  Shiraishi}, \bibinfo{person}{Akihito Iwai}, {and} \bibinfo{person}{Grigore
  Rosu}.} \bibinfo{year}{2015}\natexlab{}.
\newblock \showarticletitle{RV-Android: Efficient Parametric Android Runtime
  Verification, a Brief Tutorial}. In \bibinfo{booktitle}{\emph{Runtime
  Verification - 6th International Conference, {RV} 2015 Vienna, Austria,
  September 22-25, 2015. Proceedings}} \emph{(\bibinfo{series}{Lecture Notes in
  Computer Science})}, Vol.~\bibinfo{volume}{9333}.
  \bibinfo{publisher}{Springer}, \bibinfo{pages}{342--357}.
\newblock
\urldef\tempurl%
\url{https://doi.org/10.1007/978-3-319-23820-3_24}
\showDOI{\tempurl}


\bibitem[\protect\citeauthoryear{{Eler}, {Rojas}, {Ge}, and {Fraser}}{{Eler}
  et~al\mbox{.}}{2018}]%
        {Eler+:2018:ICST}
\bibfield{author}{\bibinfo{person}{M.~M. {Eler}}, \bibinfo{person}{J.~M.
  {Rojas}}, \bibinfo{person}{Y. {Ge}}, {and} \bibinfo{person}{G. {Fraser}}.}
  \bibinfo{year}{2018}\natexlab{}.
\newblock \showarticletitle{Automated Accessibility Testing of Mobile Apps}. In
  \bibinfo{booktitle}{\emph{2018 IEEE 11th International Conference on Software
  Testing, Verification and Validation (ICST)}}.
\newblock


\bibitem[\protect\citeauthoryear{Falcone, Currea, and Jaber}{Falcone
  et~al\mbox{.}}{2012}]%
        {Falcone+:2012:RV}
\bibfield{author}{\bibinfo{person}{Ylies Falcone}, \bibinfo{person}{Sebastian
  Currea}, {and} \bibinfo{person}{Mohamad Jaber}.}
  \bibinfo{year}{2012}\natexlab{}.
\newblock \showarticletitle{Runtime verification and enforcement for Android
  applications with RV-Droid}. In \bibinfo{booktitle}{\emph{International
  Conference on Runtime Verification}}. Springer, \bibinfo{pages}{88--95}.
\newblock


\bibitem[\protect\citeauthoryear{Ganger, Duryea, and Hu}{Ganger
  et~al\mbox{.}}{2016}]%
        {Ganger+:2016:JDAIP}
\bibfield{author}{\bibinfo{person}{Michael Ganger}, \bibinfo{person}{Ethan
  Duryea}, {and} \bibinfo{person}{Wei Hu}.} \bibinfo{year}{2016}\natexlab{}.
\newblock \showarticletitle{Double Sarsa and double expected Sarsa with shallow
  and deep learning}.
\newblock \bibinfo{journal}{\emph{Journal of Data Analysis and Information
  Processing}} \bibinfo{volume}{4}, \bibinfo{number}{04}
  (\bibinfo{year}{2016}), \bibinfo{pages}{159}.
\newblock


\bibitem[\protect\citeauthoryear{Gartner}{Gartner}{2018}]%
        {Gartner:2018}
Gartner \bibinfo{year}{2018}\natexlab{}.
\newblock \bibinfo{title}{Market Share: Final PCs, ultramobiles and mobile
  phones, all countries, 4q17 update}.
\newblock
\newblock


\bibitem[\protect\citeauthoryear{Google}{Google}{[n.d.]}]%
        {Google:MONKEY}
\bibfield{author}{\bibinfo{person}{Google}.} \bibinfo{year}{[n.d.]}\natexlab{}.
\newblock \bibinfo{title}{{Android UI/application exerciser monkey}}.
\newblock
\newblock
\newblock
\shownote{{\\http://developer.android.com/tools/help/monkey.html}.}


\bibitem[\protect\citeauthoryear{GPLAYDATA}{GPLAYDATA}{[n.d.]}]%
        {GPLAYDATA}
GPLAYDATA \bibinfo{year}{[n.d.]}\natexlab{}.
\newblock
\newblock
\newblock
\shownote{https://en.wikipedia.org/wiki/Google Play.}


\bibitem[\protect\citeauthoryear{Gultnieks}{Gultnieks}{2010}]%
        {Gultnieks:2010:FDROID}
\bibfield{author}{\bibinfo{person}{Ciaran Gultnieks}.}
  \bibinfo{year}{2010}\natexlab{}.
\newblock \bibinfo{title}{{F-Droid Benchmarks}}.
\newblock
\newblock
\newblock
\shownote{https://f-droid.org/.}


\bibitem[\protect\citeauthoryear{Hao, Liu, Nath, Halfond, and Govindan}{Hao
  et~al\mbox{.}}{2014}]%
        {Hao+:2014:MOBISYS}
\bibfield{author}{\bibinfo{person}{Shuai Hao}, \bibinfo{person}{Bin Liu},
  \bibinfo{person}{Suman Nath}, \bibinfo{person}{William~G.J. Halfond}, {and}
  \bibinfo{person}{Ramesh Govindan}.} \bibinfo{year}{2014}\natexlab{}.
\newblock \showarticletitle{{PUMA: Programmable UI-automation for Large-scale
  Dynamic Analysis of Mobile Apps}}. In \bibinfo{booktitle}{\emph{12th Annual
  International Conference on Mobile Systems, Applications, and Services
  (MobiSys)}}. \bibinfo{pages}{204--217}.
\newblock


\bibitem[\protect\citeauthoryear{Hasanbeig, Abate, and Kroening}{Hasanbeig
  et~al\mbox{.}}{2019a}]%
        {Hasanbeig+:2019:arXiv}
\bibfield{author}{\bibinfo{person}{Mohammadhosein Hasanbeig},
  \bibinfo{person}{Alessandro Abate}, {and} \bibinfo{person}{Daniel Kroening}.}
  \bibinfo{year}{2019}\natexlab{a}.
\newblock \showarticletitle{Certified Reinforcement Learning with Logic
  Guidance}.
\newblock \bibinfo{journal}{\emph{arXiv preprint arXiv:1902.00778}}
  (\bibinfo{year}{2019}).
\newblock


\bibitem[\protect\citeauthoryear{Hasanbeig, Abate, and Kroening}{Hasanbeig
  et~al\mbox{.}}{2019b}]%
        {Hasanbeig+:2019:AAMAS}
\bibfield{author}{\bibinfo{person}{Mohammadhosein Hasanbeig},
  \bibinfo{person}{Alessandro Abate}, {and} \bibinfo{person}{Daniel Kroening}.}
  \bibinfo{year}{2019}\natexlab{b}.
\newblock \showarticletitle{Logically-Constrained Neural Fitted Q-Iteration}.
  In \bibinfo{booktitle}{\emph{Proceedings of the 18th International Conference
  on Autonomous Agents and MultiAgent Systems}}. International Foundation for
  Autonomous Agents and Multiagent Systems, \bibinfo{pages}{2012--2014}.
\newblock


\bibitem[\protect\citeauthoryear{Koroglu and Sen}{Koroglu and Sen}{2018}]%
        {KorogluSen:2018:FASE}
\bibfield{author}{\bibinfo{person}{Yavuz Koroglu} {and} \bibinfo{person}{Alper
  Sen}.} \bibinfo{year}{2018}\natexlab{}.
\newblock \showarticletitle{TCM: Test Case Mutation to Improve Crash Detection
  in Android}. In \bibinfo{booktitle}{\emph{Fundamental Approaches to Software
  Engineering}}.
\newblock


\bibitem[\protect\citeauthoryear{Koroglu, Sen, Muslu, Mete, Ulker, Tanriverdi,
  and Donmez}{Koroglu et~al\mbox{.}}{2018}]%
        {Koroglu+:2018:ICST}
\bibfield{author}{\bibinfo{person}{Yavuz Koroglu}, \bibinfo{person}{Alper Sen},
  \bibinfo{person}{Ozlem Muslu}, \bibinfo{person}{Yunus Mete},
  \bibinfo{person}{Ceyda Ulker}, \bibinfo{person}{Tolga Tanriverdi}, {and}
  \bibinfo{person}{Yunus Donmez}.} \bibinfo{year}{2018}\natexlab{}.
\newblock \showarticletitle{{QBE: QLearning-Based Exploration of Android
  Applications}}. In \bibinfo{booktitle}{\emph{IEEE International Conference on
  Software Testing, Verification and Validation (ICST)}}.
\newblock


\bibitem[\protect\citeauthoryear{Laud}{Laud}{2004}]%
        {Laud:2004}
\bibfield{author}{\bibinfo{person}{Adam~Daniel Laud}.}
  \bibinfo{year}{2004}\natexlab{}.
\newblock \bibinfo{booktitle}{\emph{Theory and application of reward shaping in
  reinforcement learning}}.
\newblock \bibinfo{type}{{T}echnical {R}eport}.
\newblock


\bibitem[\protect\citeauthoryear{Li, Yang, Guo, and Chen}{Li
  et~al\mbox{.}}{2017}]%
        {Li+:2017:ICSE-C}
\bibfield{author}{\bibinfo{person}{Yuanchun Li}, \bibinfo{person}{Ziyue Yang},
  \bibinfo{person}{Yao Guo}, {and} \bibinfo{person}{Xiangqun Chen}.}
  \bibinfo{year}{2017}\natexlab{}.
\newblock \showarticletitle{DroidBot: a lightweight UI-guided test input
  generator for Android}. In \bibinfo{booktitle}{\emph{2017 IEEE/ACM 39th
  International Conference on Software Engineering Companion (ICSE-C)}}.
\newblock
\newblock
\shownote{{https://github.com/honeynet/droidbot}.}


\bibitem[\protect\citeauthoryear{Linares-V\'{a}squez, White,
  Bernal-C\'{a}rdenas, Moran, and Poshyvanyk}{Linares-V\'{a}squez
  et~al\mbox{.}}{2015}]%
        {Linares-Vasquez+:2015:MSR}
\bibfield{author}{\bibinfo{person}{Mario Linares-V\'{a}squez},
  \bibinfo{person}{Martin White}, \bibinfo{person}{Carlos Bernal-C\'{a}rdenas},
  \bibinfo{person}{Kevin Moran}, {and} \bibinfo{person}{Denys Poshyvanyk}.}
  \bibinfo{year}{2015}\natexlab{}.
\newblock \showarticletitle{{Mining Android App Usages for Generating
  Actionable GUI-based Execution Scenarios}}. In \bibinfo{booktitle}{\emph{12th
  Working Conference on Mining Software Repositories (MSR)}}.
  \bibinfo{pages}{111--122}.
\newblock


\bibitem[\protect\citeauthoryear{Liu, Xu, Cheung, and Lu}{Liu
  et~al\mbox{.}}{2014}]%
        {Liu+:2014:IEEE}
\bibfield{author}{\bibinfo{person}{Yepang Liu}, \bibinfo{person}{Chang Xu},
  \bibinfo{person}{Shing-Chi Cheung}, {and} \bibinfo{person}{Jian Lu}.}
  \bibinfo{year}{2014}\natexlab{}.
\newblock \showarticletitle{GreenDroid: Automated Diagnosis of Energy
  Inefficiency for Smartphone Applications}.
\newblock \bibinfo{journal}{\emph{IEEE Transactions on Software Engineering}}
  (\bibinfo{year}{2014}).
\newblock


\bibitem[\protect\citeauthoryear{Machiry, Tahiliani, and Naik}{Machiry
  et~al\mbox{.}}{2013}]%
        {Machiry+:2013:FSE}
\bibfield{author}{\bibinfo{person}{Aravind Machiry}, \bibinfo{person}{Rohan
  Tahiliani}, {and} \bibinfo{person}{Mayur Naik}.}
  \bibinfo{year}{2013}\natexlab{}.
\newblock \showarticletitle{{Dynodroid: An Input Generation System for Android
  Apps}}. In \bibinfo{booktitle}{\emph{9th Joint Meeting on Foundations of
  Software Engineering (ESEC/FSE)}}.
\newblock
\newblock
\shownote{{https://dynodroid.github.io}.}


\bibitem[\protect\citeauthoryear{Mahmood, Mirzaei, and Malek}{Mahmood
  et~al\mbox{.}}{2014}]%
        {Mahmood+:2014:FSE}
\bibfield{author}{\bibinfo{person}{Riyadh Mahmood}, \bibinfo{person}{Nariman
  Mirzaei}, {and} \bibinfo{person}{Sam Malek}.}
  \bibinfo{year}{2014}\natexlab{}.
\newblock \showarticletitle{{EvoDroid: Segmented Evolutionary Testing of
  Android Apps}}. In \bibinfo{booktitle}{\emph{22Nd ACM SIGSOFT International
  Symposium on Foundations of Software Engineering (FSE)}}.
  \bibinfo{pages}{599--609}.
\newblock


\bibitem[\protect\citeauthoryear{Mao, Alizadeh, Menache, and Kandula}{Mao
  et~al\mbox{.}}{2016a}]%
        {Mao+:ACM:2016}
\bibfield{author}{\bibinfo{person}{Hongzi Mao}, \bibinfo{person}{Mohammad
  Alizadeh}, \bibinfo{person}{Ishai Menache}, {and} \bibinfo{person}{Srikanth
  Kandula}.} \bibinfo{year}{2016}\natexlab{a}.
\newblock \showarticletitle{Resource management with deep reinforcement
  learning}. In \bibinfo{booktitle}{\emph{Proceedings of the 15th ACM Workshop
  on Hot Topics in Networks}}. ACM, \bibinfo{pages}{50--56}.
\newblock


\bibitem[\protect\citeauthoryear{Mao, Harman, and Jia}{Mao
  et~al\mbox{.}}{2016b}]%
        {Mao+:2016:ISSTA}
\bibfield{author}{\bibinfo{person}{Ke Mao}, \bibinfo{person}{Mark Harman},
  {and} \bibinfo{person}{Yue Jia}.} \bibinfo{year}{2016}\natexlab{b}.
\newblock \showarticletitle{{Sapienz: Multi-objective Automated Testing for
  Android Applications}}. In \bibinfo{booktitle}{\emph{25th International
  Symposium on Software Testing and Analysis (ISSTA)}}.
  \bibinfo{pages}{94--105}.
\newblock


\bibitem[\protect\citeauthoryear{Mariani, Pezze, Riganelli, and
  Santoro}{Mariani et~al\mbox{.}}{2012}]%
        {Mariani+:2012:ICST}
\bibfield{author}{\bibinfo{person}{Leonardo Mariani}, \bibinfo{person}{Mauro
  Pezze}, \bibinfo{person}{Oliviero Riganelli}, {and} \bibinfo{person}{Mauro
  Santoro}.} \bibinfo{year}{2012}\natexlab{}.
\newblock \showarticletitle{Autoblacktest: Automatic black-box testing of
  interactive applications}. In \bibinfo{booktitle}{\emph{2012 IEEE Fifth
  International Conference on Software Testing, Verification and Validation}}.
  IEEE, \bibinfo{pages}{81--90}.
\newblock


\bibitem[\protect\citeauthoryear{Mirzaei, Garcia, Bagheri, Sadeghi, and
  Malek}{Mirzaei et~al\mbox{.}}{2016}]%
        {Mirzaei+:2016:ICSE}
\bibfield{author}{\bibinfo{person}{Nariman Mirzaei}, \bibinfo{person}{Joshua
  Garcia}, \bibinfo{person}{Hamid Bagheri}, \bibinfo{person}{Alireza Sadeghi},
  {and} \bibinfo{person}{Sam Malek}.} \bibinfo{year}{2016}\natexlab{}.
\newblock \showarticletitle{Reducing combinatorics in GUI testing of android
  applications}. In \bibinfo{booktitle}{\emph{2016 IEEE/ACM 38th International
  Conference on Software Engineering (ICSE)}}. IEEE, \bibinfo{pages}{559--570}.
\newblock


\bibitem[\protect\citeauthoryear{Mnih, Kavukcuoglu, Silver, Graves, Antonoglou,
  Wierstra, and Riedmiller}{Mnih et~al\mbox{.}}{2013}]%
        {Mnih+:2013:arXiv}
\bibfield{author}{\bibinfo{person}{Volodymyr Mnih}, \bibinfo{person}{Koray
  Kavukcuoglu}, \bibinfo{person}{David Silver}, \bibinfo{person}{Alex Graves},
  \bibinfo{person}{Ioannis Antonoglou}, \bibinfo{person}{Daan Wierstra}, {and}
  \bibinfo{person}{Martin Riedmiller}.} \bibinfo{year}{2013}\natexlab{}.
\newblock \showarticletitle{Playing atari with deep reinforcement learning}.
\newblock \bibinfo{journal}{\emph{arXiv preprint arXiv:1312.5602}}
  (\bibinfo{year}{2013}).
\newblock


\bibitem[\protect\citeauthoryear{Moran, V{\'{a}}squez, Bernal{-}C{\'{a}}rdenas,
  Vendome, and Poshyvanyk}{Moran et~al\mbox{.}}{2016}]%
        {Moran+:2016:ICST}
\bibfield{author}{\bibinfo{person}{Kevin Moran}, \bibinfo{person}{Mario~Linares
  V{\'{a}}squez}, \bibinfo{person}{Carlos Bernal{-}C{\'{a}}rdenas},
  \bibinfo{person}{Christopher Vendome}, {and} \bibinfo{person}{Denys
  Poshyvanyk}.} \bibinfo{year}{2016}\natexlab{}.
\newblock \showarticletitle{{Automatically Discovering, Reporting and
  Reproducing Android Application Crashes}}. In
  \bibinfo{booktitle}{\emph{{IEEE} International Conference on Software
  Testing, Verification and Validation (ICST)}}. \bibinfo{pages}{33--44}.
\newblock
\newblock
\shownote{{https://www.android-dev-tools.com/crashscope-home}.}


\bibitem[\protect\citeauthoryear{Piejko}{Piejko}{2016}]%
        {Piejko:2016:DEVICEATLAS}
\bibfield{author}{\bibinfo{person}{Pawel Piejko}.}
  \bibinfo{year}{2016}\natexlab{}.
\newblock \bibinfo{title}{{16 mobile market statistics you should know in
  20\-16}}.
\newblock
\newblock
\newblock
\shownote{https://deviceatlas.com/blog/16-mobile-market-statistics-you-should-know-2016.}


\bibitem[\protect\citeauthoryear{Silver, Hubert, Schrittwieser, Antonoglou,
  Lai, Guez, Lanctot, Sifre, Kumaran, Graepel, et~al\mbox{.}}{Silver
  et~al\mbox{.}}{2017}]%
        {Silver+:2017:arXiv}
\bibfield{author}{\bibinfo{person}{David Silver}, \bibinfo{person}{Thomas
  Hubert}, \bibinfo{person}{Julian Schrittwieser}, \bibinfo{person}{Ioannis
  Antonoglou}, \bibinfo{person}{Matthew Lai}, \bibinfo{person}{Arthur Guez},
  \bibinfo{person}{Marc Lanctot}, \bibinfo{person}{Laurent Sifre},
  \bibinfo{person}{Dharshan Kumaran}, \bibinfo{person}{Thore Graepel},
  {et~al\mbox{.}}} \bibinfo{year}{2017}\natexlab{}.
\newblock \showarticletitle{Mastering chess and shogi by self-play with a
  general reinforcement learning algorithm}.
\newblock \bibinfo{journal}{\emph{arXiv preprint arXiv:1712.01815}}
  (\bibinfo{year}{2017}).
\newblock


\bibitem[\protect\citeauthoryear{Su, Meng, Chen, Wu, Yang, Yao, Pu, Liu, and
  Su}{Su et~al\mbox{.}}{2017}]%
        {Su+:2017:FSE}
\bibfield{author}{\bibinfo{person}{Ting Su}, \bibinfo{person}{Guozhu Meng},
  \bibinfo{person}{Yuting Chen}, \bibinfo{person}{Ke Wu},
  \bibinfo{person}{Weiming Yang}, \bibinfo{person}{Yao Yao},
  \bibinfo{person}{Geguang Pu}, \bibinfo{person}{Yang Liu}, {and}
  \bibinfo{person}{Zhendong Su}.} \bibinfo{year}{2017}\natexlab{}.
\newblock \showarticletitle{Guided, stochastic model-based GUI testing of
  Android apps}. In \bibinfo{booktitle}{\emph{Proceedings of the 2017 11th
  Joint Meeting on Foundations of Software Engineering}}.
\newblock
\newblock
\shownote{{https://tingsu.github.io/files/stoat.html}.}


\bibitem[\protect\citeauthoryear{Sun, Rosa, Javed, and Binder}{Sun
  et~al\mbox{.}}{2017}]%
        {Sun+:2017:ICST}
\bibfield{author}{\bibinfo{person}{Haiyang Sun}, \bibinfo{person}{Andrea Rosa},
  \bibinfo{person}{Omar Javed}, {and} \bibinfo{person}{Walter Binder}.}
  \bibinfo{year}{2017}\natexlab{}.
\newblock \showarticletitle{ADRENALIN-RV: android runtime verification using
  load-time weaving}. In \bibinfo{booktitle}{\emph{2017 IEEE International
  Conference on Software Testing, Verification and Validation (ICST)}}. IEEE,
  \bibinfo{pages}{532--539}.
\newblock


\bibitem[\protect\citeauthoryear{Sutton and Barto}{Sutton and Barto}{1998}]%
        {SuttonBarto:1998:RL}
\bibfield{author}{\bibinfo{person}{Richard~S. Sutton} {and}
  \bibinfo{person}{Andrew~G. Barto}.} \bibinfo{year}{1998}\natexlab{}.
\newblock \bibinfo{booktitle}{\emph{Introduction to Reinforcement Learning}
  (\bibinfo{edition}{1st} ed.)}.
\newblock \bibinfo{publisher}{MIT Press}, \bibinfo{address}{Cambridge, MA,
  USA}.
\newblock
\showISBNx{0262193981}


\bibitem[\protect\citeauthoryear{Toro~Icarte, Klassen, Valenzano, and
  McIlraith}{Toro~Icarte et~al\mbox{.}}{2018}]%
        {Icarte+:2018:AAMAS}
\bibfield{author}{\bibinfo{person}{Rodrigo Toro~Icarte},
  \bibinfo{person}{Toryn~Q. Klassen}, \bibinfo{person}{Richard Valenzano},
  {and} \bibinfo{person}{Sheila~A. McIlraith}.}
  \bibinfo{year}{2018}\natexlab{}.
\newblock \showarticletitle{Teaching Multiple Tasks to an RL Agent Using LTL}.
  In \bibinfo{booktitle}{\emph{Proceedings of the 17th International Conference
  on Autonomous Agents and MultiAgent Systems}} \emph{(\bibinfo{series}{AAMAS
  '18})}. \bibinfo{publisher}{International Foundation for Autonomous Agents
  and Multiagent Systems}, \bibinfo{address}{Richland, SC},
  \bibinfo{pages}{452--461}.
\newblock
\urldef\tempurl%
\url{http://dl.acm.org/citation.cfm?id=3237383.3237452}
\showURL{%
\tempurl}


\bibitem[\protect\citeauthoryear{Wen, Ehlers, and Topcu}{Wen
  et~al\mbox{.}}{2015}]%
        {Wen+:2015:IROS}
\bibfield{author}{\bibinfo{person}{Min Wen}, \bibinfo{person}{R{\"u}diger
  Ehlers}, {and} \bibinfo{person}{Ufuk Topcu}.}
  \bibinfo{year}{2015}\natexlab{}.
\newblock \showarticletitle{Correct-by-synthesis reinforcement learning with
  temporal logic constraints}. In \bibinfo{booktitle}{\emph{IEEE/RSJ
  International Conference on Intelligent Robots and Systems (IROS)}}.
\newblock


\bibitem[\protect\citeauthoryear{Yan, Pan, Li, Yan, and Zhang}{Yan
  et~al\mbox{.}}{2018}]%
        {Yan+:2018:ISSTA}
\bibfield{author}{\bibinfo{person}{Jiwei Yan}, \bibinfo{person}{Linjie Pan},
  \bibinfo{person}{Yaqi Li}, \bibinfo{person}{Jun Yan}, {and}
  \bibinfo{person}{Jian Zhang}.} \bibinfo{year}{2018}\natexlab{}.
\newblock \showarticletitle{LAND: A User-friendly and Customizable Test
  Generation Tool for Android Apps}. In \bibinfo{booktitle}{\emph{Proceedings
  of the 27th ACM SIGSOFT International Symposium on Software Testing and
  Analysis (ISSTA)}}.
\newblock


\bibitem[\protect\citeauthoryear{Yang, Prasad, and Xie}{Yang
  et~al\mbox{.}}{2013}]%
        {Yang+:2013:FASE}
\bibfield{author}{\bibinfo{person}{Wei Yang}, \bibinfo{person}{Mukul~R.
  Prasad}, {and} \bibinfo{person}{Tao Xie}.} \bibinfo{year}{2013}\natexlab{}.
\newblock \showarticletitle{{A Grey-box Approach for Automated GUI-model
  Generation of Mobile Applications}}. In \bibinfo{booktitle}{\emph{16th
  International Conference on Fundamental Approaches to Software Engineering
  (FASE)}}. \bibinfo{pages}{250--265}.
\newblock


\bibitem[\protect\citeauthoryear{Zaeem, Prasad, and Khurshid}{Zaeem
  et~al\mbox{.}}{2014}]%
        {Zaeem+:2014:ICST}
\bibfield{author}{\bibinfo{person}{Razieh~Nokhbeh Zaeem},
  \bibinfo{person}{Mukul~R. Prasad}, {and} \bibinfo{person}{Sarfraz Khurshid}.}
  \bibinfo{year}{2014}\natexlab{}.
\newblock \showarticletitle{{Automated Generation of Oracles for Testing
  User-Interaction Features of Mobile Apps}}. In \bibinfo{booktitle}{\emph{IEEE
  International Conference on Software Testing, Verification, and Validation
  (ICST)}}.
\newblock


\bibitem[\protect\citeauthoryear{Zhou, Li, and Zare}{Zhou
  et~al\mbox{.}}{2017}]%
        {Zhou+:ACS:2017}
\bibfield{author}{\bibinfo{person}{Zhenpeng Zhou}, \bibinfo{person}{Xiaocheng
  Li}, {and} \bibinfo{person}{Richard~N Zare}.}
  \bibinfo{year}{2017}\natexlab{}.
\newblock \showarticletitle{Optimizing chemical reactions with deep
  reinforcement learning}.
\newblock \bibinfo{journal}{\emph{ACS central science}} \bibinfo{volume}{3},
  \bibinfo{number}{12} (\bibinfo{year}{2017}), \bibinfo{pages}{1337--1344}.
\newblock


\end{thebibliography}

\end{document}